\documentclass[12pt,preprint]{aastex}

\shorttitle{}
\shortauthors{Nesvorn\'y et al.}

\begin{document}
\baselineskip 19.pt

\title{OSSOS XX: The Meaning of Kuiper Belt Colors}

\author{David Nesvorn\'y$^1$, David Vokrouhlick\'y$^2$,
Mike Alexandersen$^{3}$,
Michele T. Bannister$^4$,
Laura E. Buchanan$^4$, %ORCID 0000-0002-8032-4528
Ying-Tung Chen$^3$,
Brett J. Gladman$^5$,
Stephen D. J. Gwyn$^{6}$,
J. J. Kavelaars$^{6,7}$,
Jean-Marc Petit$^{8}$,
Megan E. Schwamb$^4$, %ORCID 0000-0003-4365-1455
Kathryn Volk$^{9,*}$} 

\affil{(1) Department of Space Studies, Southwest Research Institute,\\
1050 Walnut St., Suite 300, Boulder, CO, 80302, United States}
\affil{(2) Institute of Astronomy, Charles University,\\ 
V Hole\v{s}ovi\v{c}k\'ach 2, CZ--18000 Prague 8, Czech Republic}
\affil{(3) Institute of Astronomy and Astrophysics, Academia Sinica,\\ 11F of AS/NTU Astronomy-Mathematics Building, 
Nr. 1 Roosevelt Rd., Sec. 4, Taipei 10617, Taiwan, R.O.C.}
\affil{(4) Astrophysics Research Centre, Queen's University Belfast,\\ Belfast BT7 1NN, United Kingdom}
\affil{(5) Department of Physics and Astronomy, University of British Columbia,\\ Vancouver, BC, Canada}
\affil{(6) NRC-Herzberg Astronomy and Astrophysics, National Research Council of Canada,\\ 5071 West Saanich Rd, 
Victoria, British Columbia V9E 2E7, Canada}
\affil{(7) Department of Physics and Astronomy, University of Victoria,\\ Elliott Building, 3800 Finnerty Rd, 
Victoria, BC V8P 5C2, Canada}
\affil{(8) Institut UTINAM UMR6213, CNRS, Univ. Bourgogne Franche-Comt\'e,\\ OSU Theta F25000 Besan\c{c}on, France} 
\affil{(9) Lunar and Planetary Laboratory, University of Arizona,\\ 1629 E University Blvd, Tucson, AZ 85721, United States}
\affil{* The OSSOS team members are listed in alphabetical order}

\begin{abstract} 
Observations show that 100-km-class Kuiper belt objects (KBOs) can be divided in (at least) two 
color groups, hereafter red (R; $g-i<1.2$) and very red (VR; $g-i>1.2$), reflecting a difference in their surface 
composition. This is thought to imply that KBOs formed over a relatively wide range of radial distance, $r$. 
The cold classicals at $42<r<47$ au are predominantly VR and known Neptune Trojans at 
$r \simeq 30$~au are mostly R. Intriguingly, however, the dynamically hot KBOs show a mix 
of R and VR colors and no correlation of color with $r$. Here we perform migration/instability 
simulations where the Kuiper belt is populated from an extended planetesimal disk. We find that the 
color observations can be best understood if R objects formed at $r<r^*$ and VR objects at $r>r^*$, 
with $30<r^*<40$ au. The proposed transition at $30<r^*<40$~au would explain why the VR objects in 
the dynamically hot population have smaller orbital inclinations than the R objects, because the 
orbital excitation from Neptune weakens for orbits starting beyond 30 au. Possible causes of the 
R-VR color bimodality are discussed.
\end{abstract}


\section{Introduction}

The vast majority of KBOs are too faint for spectroscopic observations, but their surface composition 
can be studied with broadband photometry. Photometric observations indicate that 
the color distribution of KBOs is bimodal\footnote{In fact, the color distribution of KBOs is complex
and many color subdivisions exist (e.g., Pike et al. 2017). We do not discuss these color sub-groups 
here because our work does not offer any new insight into subtle compositional differences. In addition,
note that different terminologies are currently in use. For example, the VR color, as used here, is 
often referred to as ``ultra-red'' (e.g., Sheppard 2012, Peixinho et al. 2015).} with red (R; defined 
as $g-i<1.2$ in Wong \& Brown (2017); observations made in the $ugriz$ magnitude system) and very red 
(VR; $g-i>1.2$) classes (e.g., Luu \& Jewitt 1996, 1998; Tegler \& Romanishin 1998, 2000; Jewitt \& Luu 1998, 
2001; Peixinho et al. 2003, 2008, 2012, 2015; Tegler et al. 2003, 2016; Barucci et al. 2005; Sheppard 2012; 
Fraser \& Brown 2012; Wong \& Brown 2017; Pike et al. 2017; Jewitt 2018; Schwamb et al. 2019, Marsset et al. 2019). 
Known Neptune Trojans at radial distance $r \simeq 30$ au are R (Sheppard \& Trujillo 2006, Parker et al. 2013, 
Jewitt 2018), with one known exception (2013 VX30; Lin et al. 2019), and most classified cold 
classicals (CCs)\footnote{Based on their orbits, KBOs can be classified into several categories: 
classical Kuiper belt (CKB), resonant populations, scattered disk objects (SDOs), etc. (Gladman et 
al. 2008). Most known KBOs reside in the main CKB, which is located between the 3:2 and 2:1 resonances 
with Neptune ($39.4<r<47.8$ au). It is furthermore useful to divide the CKB into dynamically ``cold'' 
(CCs; orbital inclinations $i<5^\circ$) and ``hot'' components (HCs; $i>5^\circ$), mainly because the 
inclination distribution in the CKB is bimodal (Brown 2001) and CCs have unique physical properties 
(e.g., VR colors, Jewitt \& Luu 1998, Tegler \& Romanishin 2000; large binary fraction, Noll et al. 2020).
In this text, the dynamically hot KBO population is defined an ensemble of HCs, resonant populations and 
SDOs, whereas the dynamically cold population is the same as CCs.} 
with semimajor axes $42<a<47$~au are VR. This has been taken as evidence that colors have something to do 
with the distance at which different objects formed. Confusing matters, however, the dynamically hot 
populations with $30<a<50$ au show a mix of R and VR colors, and there does not appear to be any 
obvious correlation of colors with~$r$ (e.g., Peixinho et al. 2015, Marsset et al. 2019).

Brown et al. (2011) proposed that the early surface compositions of KBOs were set by volatile evaporation
after the objects formed. A strong gradient in surface composition, coupled with UV irradiation and 
particle impacts, then presumably led to the surface colors that we see today. For example, the sublimation 
line of (pure) ammonia, NH$_3$, is near 34 au (Brown et al. 2011). Objects formed at the current 
location of CCs may therefore uniquely retain NH$_3$, which has been shown to affect irradiation chemistry 
and could plausibly lead to the VR colors of these objects. {\it But how to interpret the R colors of Neptune 
Trojans and the bimodal distribution of colors in the hot population?}

Neptune Trojans were presumably trapped as co-orbitals during Neptune's migration (e.g., Nesvorn\'y 
\& Vokrouhlick\'y  2009, Parker 2015, Gomes \& Nesvorn\'y 2016). Their inferred formation location
is $r \simeq 25$-30 au. The predominantly R colors of Neptune Trojans (Jewitt 2018, Lin et al. 2019) would 
thus be hard to understand if the R to VR transition is related the sublimation line of the hydrogen sulfide ice 
(H$_2$S, $r\simeq15$-20 au), as suggested by Wong \& Brown (2016, 2017). Instead, the R colors of 
Neptune Trojans seem to imply that the transition occurred farther out, probably beyond $\sim$30 au. 
This reasoning leads to an impasse, however, because our best dynamical models suggest that the 
dynamically hot KBOs were implanted onto their current orbits from the massive planetesimal disk 
inside of $30$ au. Their colors should thus be uniformly R, just like Neptune Trojans, but they are not. 

\section{Color Hypothesis}

Here we examine the possibility that the hot populations (i.e., HCs, Plutinos\footnote{Plutinos in the 
3:2 resonance with Neptune are the most populated and best characterized resonant population. Here we 
focus on this population. Other resonant populations will be considered in future work.}, SDOs) 
in the present-day Kuiper belt are a mix of bodies implanted from the 
massive disk {\it below} 30~au (source of R) and the low-mass disk extension {\it beyond} 30~au 
(source of VR). On one hand, the surface density of planetesimals must have been quite low at $r>30$ au 
for Neptune to stop at 30~au (Gomes et al. 2004, Nesvorn\'y 2018). The outer disk extension thus 
represents a smaller source reservoir than the massive disk below 30~au. On the other hand, the 
chances to evolve from 30-40~au onto a dynamically hot orbit in the Kuiper belt are better (e.g., 
Hahn \& Malhotra 2005). It is thus plausible that a good share of hot KBOs come from the 
30-40 au region.\footnote{Note that planetesimals starting beyond 40~au remain on low inclination orbits 
during Neptune's migration (e.g., Batygin et al. 2011, Nesvorn\'y 2015b); the $r>40$~au region is 
therefore not a major source of dynamically hot KBOs.}

The proposed color transition at $r^*>30$ au could explain why the VR objects in the dynamically hot 
population have smaller orbital inclinations than the R objects (e.g., Tegler \& Romanishin 2000; 
Trujillo \& Brown 2002; Hainaut \& Delsanti 2002; Peixinho et al. 2008, 2015; Marsset et al. 2019), 
because the orbital excitation from Neptune is expected to weaken for orbits starting beyond 30 au. 
We investigate this issue in detail in Section 4.3. In contrast, no such correlation would be expected 
if both the R and VR objects started below 30 au, where Neptune's gravitational effects are uniformly 
strong (Nesvorn\'y 2015a). 

If some of the R objects can be pushed out from $r<r^*$ into the CC population, this could explain 
the ``blue'' CC binaries reported in Fraser et al. (2017) and indicate that $r^*>35$ au. Note that 
the analysis presented here aims at explaining the global distribution of KBO colors, including 
the color-inclination correlation; this is well beyond the scope of the analysis of blue binaries in 
Fraser et al. (2017). Our color hypothesis would also be consistent with the R colors of Jupiter 
Trojans (Emery et al. 2015) and irregular satellites of the giant planets (Graykowski \& Jewitt~2018), 
because they are thought to be captured from the massive disk below 30 au, and thus expected to be R.

Pike et al. (2017) pointed out that cold and hot KBOs with $g-r>0.8$ (roughly the VR 
category here) have different $r-z$ colors (hot VR KBOs exhibit redder $r-z$ colors).
This was used in Schwamb et al. (2019) to propose that the original planetesimal disk had 
two color transitions: one at $\sim$33 au, from VR with redder $r-z$ colors to R
(called ``neutral'' in Schwamb et al.) and another one at $\sim$39 au, from R to VR with 
bluer $r-z$ colors. This cannot work, however, because: (i) it would not fit the predominantly
R colors of Neptune Trojans (Jewitt 2018, Lin et al. 2019), and (ii) VR objects starting below 33 au would 
end up on orbits with {\it higher} orbital inclinations than R objects starting at 33-39~au (Section 4.3), 
which is opposite to what the color observations indicate (e.g., Marsset et al. 2019). Pike et al.'s result 
is more likely related to a change of $r-z$ with original orbital radius from $r^*$ to the location 
of CCs at $r>42$ au. 
 
\section{Methods}

The occurrence of R and VR objects in each KBO category can be determined, in the context of the
suggested model, from the initial disk profile (massive at $r<30$ au with decreasing surface density 
beyond 30 au), the radial distance $r^*$ which marked the original transition from R to VR colors, 
and the implantation probability from $r$ to a specific dynamical class. This is the main goal of 
the work presented here. We aim at identifying the disk profiles and the range of $r^*$ values that 
best fit the existing color data. 

\subsection{Integration Method} 

The numerical integrations conducted here consist of tracking the orbits of four giant planets 
(Jupiter to Neptune) and a large number of particles ($2\times10^6$) representing the original 
trans-Neptunian disk. To set up an integration, Uranus and Neptune are placed inside of their current orbits and  
migrated outward. The {\tt swift\_rmvs4} code, part of the {\it Swift} $N$-body integration package 
(Levison \& Duncan 1994), is used to follow the orbits of planets and massless disk particles. 
The code was modified to include additional forces that mimic the radial migration and damping of 
planetary orbits. These forces are parametrized by the exponential e-folding timescale, $\tau$.

\begin{figure}
\epsscale{0.5}
\plotone{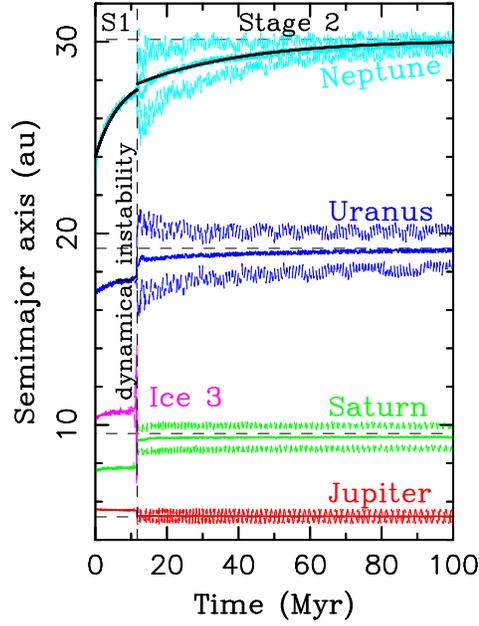}
\caption{The orbit histories of the giant planets obtained in a self-consistent migration simulation 
with a planetesimal disk between 24 and 60 au (total disk mass $M_{\rm disk}=20$~$M_{\rm Earth}$). The initial 
surface density of planetesimals was assumed to follow a radial profile with exponentially decreasing 
surface density from 24 to 60 au with one e-fold $\Delta r = 2.5$ au ({\it no disk truncation at 30 au}). 
The plot shows the semimajor axes (solid lines) and perihelion/aphelion distances (thin dashed lines) 
of each planet's orbit. The fifth planet (labeled Ice 3) was ejected from the Solar System by Jupiter during 
the instability (integration time $t=11.7$~Myr). The solid black lines are Neptune's exponential 
migration fits with $\tau_1=12$~Myr for $t<11.7$~Myr (Stage 1; labeled S1) and $\tau_2=27$ Myr for 
$t>11.7$~Myr (Stage~2). The final orbits of the planets are a good match to those in the present Solar 
System.}
\label{case40}
\end{figure}

The migration histories of planets are informed by our best models of planetary migration/instability 
(Nesvorn\'y \& Morbidelli 2012, hereafter NM12; also see Deienno et al. 2017). In the NM12 models, 
Neptune's migration can be divided into two stages separated by a brief episode of dynamical 
instability (jumping Neptune model, Figure \ref{case40}). Neptune migrates on a circular orbit
before the instability (Stage 1).  Its eccentricity becomes excited during the instability and is 
subsequently damped by a gravitational interaction with disk planetesimals (Stage 2). The instability 
is needed, among other things (e.g., orbital eccentricity of Jupiter, asteroid belt constraints; 
Nesvorn\'y 2018), to explain the Kuiper belt kernel near 44~au (Section 4.2; Petit et al. 2011, Nesvorn\'y 2015b, 
Bannister et al. 2018). 

The orbital behavior of Neptune during the first and second migration stages can be approximated by 
$\tau_1\simeq5$-30 Myr and $\tau_2\simeq30$-100~Myr. We find that Neptune's migration in a power-law 
radial disk profile is often too fast ($\tau_1 < 10$ Myr) to satisfy the inclination constraint and 
it is difficult to fine-tune the total disk mass, $M_{\rm disk}$, to obtain $\tau_1 \gtrsim 10$~Myr. 
The power-law disks also efficiently damp Neptune's eccentricity during Stage 2, which effects the 
ability of Neptune's resonances to implant bodies on the high-inclination orbits in the Kuiper belt (Volk 
\& Malhotra 2019). The exponential disks show more promising results (e.g., $\tau_1=12$ Myr and 
$\tau_2=27$ Myr in Fig. \ref{case40}). We account for the jitter that Neptune's orbit 
experiences due to close encounters with Pluto-class objects. Neptune's grainy migration is important 
to produce the right proportion of resonant and non-resonant KBOs (Nesvorn\'y \& Vokrouhlick\'y 2016). 

All migration simulations are run to 0.5 Gyr. They are extended to 4.5 Gyr with the standard 
{\tt swift\_rmvs4} code (i.e., without migration/damping after 0.5 Gyr). We perform four new 
simulations in total (Table 1). In the first case, we adopt $\tau_1=10$ Myr and $\tau_2=30$~Myr  
as indicated by Fig. \ref{case40}. This case roughly corresponds to the shortest migration timescale that 
is required to satisfy the inclination constraint (Nesvorn\'y 2015a; but see Volk \& Malhotra 2019). In the second case, we use 
longer timescales: $\tau_1=30$ Myr and $\tau_2=100$ Myr. This case would correspond to a slower 
migration driven by a lower mass planetesimal disk. A very slow migration of Neptune during the second 
stage would be needed, for example, to explain Saturn's obliquity (Ward \& Hamilton 2004, Hamilton 
\& Ward 2004, Vokrouhlick\'y \& Nesvorn\'y 2015). These two cases bracket the interesting range of 
possibilities. We perform two simulations\footnote{A full-scale simulation with two million disk particles 
over 4.5 Gyr requires $\simeq$1000 hours on 2000 Ivy Bridge cores of the NASA Pleiades Supercomputer.}
in each case, one with Neptune's jump at the transition from Stages 1 to 2 (Nesvorn\'y 2015b) 
and one without it. The jump is implemented as an instantaneous increase of Neptune's semimajor 
axis by 0.4-0.5 au (Table 1). In the following text, the simulations without jump are labeled s10/30 and 
s30/100; the ones with jump are s10/30j and s30/100j. 

The final orbits of Uranus and Neptune are fine-tuned in these simulations such as the period ratio 
$P_{\rm N}/P_{\rm U}=1.92$-1.95, where $P_{\rm U}$ and $P_{\rm N}$ are the orbital periods of Uranus and 
Neptune (Table 1). The orbital ratio of the real planets in the current Solar System is $P_{\rm N}/P_{\rm U} =
1.96$. We opt for having model $P_{\rm N}/P_{\rm U}$ a tiny bit smaller than 1.96 to make sure that Uranus 
and Neptune are never too close to the 2:1 resonance (the effects of the 2:1 resonance can disrupt
the Kuiper belt structure, change stability of resonant populations, etc.). We also make sure that 
the final semimajor axes, eccentricities and inclinations of planets are as close to the real values 
as possible. Neptune's orbital eccentricity is assumed to increase to $e_{\rm N}=0.1$ at the transition 
from Stage 1 to Stage 2, at often seen in our self-consistent instability simulations (Fig. \ref{case40}). 
The damping routines are tuned such that the simulated orbit of Neptune ends with just the 
right orbital inclination and eccentricity (current mean $e_{\rm N}=0.009$). The cases with larger final 
$e_{\rm N}$, such as the ones described as Case A in Volk \& Malhotra (2019), are not investigated here.

\subsection{Planetesimal Disk}

Previous studies of planetary migration/instability often adopted a two-part disk structure with a 
massive planetesimal disk on the inside, a low-mass disk on the outside, and a sharp transition from 
high to low surface densities near 30 au. The inner part of the planetesimal disk, from just outside 
Neptune's initial orbit to $\sim$30~au, was estimated mass to be $M_{\rm disk}\simeq15$-20~$M_{\rm Earth}$ (NM12, Nesvorn\'y et al. 2013, 2019; 
Deienno et al. 2017). It was argued that the massive disk must have been truncated at $\sim$30~au for 
Neptune to stop at 30~au (e.g., Gomes et al. 2004). Our tests show that the disk truncation is not 
required (Fig. \ref{case40}). In fact, Neptune may have ended at $\simeq$30 au just because the 
planetesimal surface density at $\gtrsim$30 au was {\it subcritical} (Neptune stops if the density is below 
1-1.5~$M_{\rm Earth}$ au$^{-1}$; Nesvorn\'y 2018). In other words, Neptune's current orbit does not constrain 
the radial gradient of the planetesimal surface density near 30 au. Instead, it just tells us that 
the surface density was low beyond 30 au. 

Another constraint on the surface density profile beyond 30 au can be inferred from the CC population.
The mass of CCs was estimated to be $M_{\rm CC} \sim 3 \times 10^{-4}$ $M_{\rm Earth}$ in Fraser et al. (2014) but we find 
$M_{\rm CC}=(3 \pm 2) \times 10^{-3}$ $M_{\rm Earth}$ from OSSOS (Outer Solar System Origins Survey; Appendix A). 
The difference is caused, at least in part, by different observational datasets used for the analyses (OSSOS biases are carefully characterized) and 
different magnitude distribution assumptions (we use an exponentially tampered size distribution in 
Appendix A). Nesvorn\'y (2015b) found that this represents only a fraction of the original population of 
planetesimals at 45 au. Adopting our OSSOS estimate, we can therefore very roughly estimate that the original 
disk mass density at 45 au was $\sim 10^{-3}$ $M_{\rm Earth}$ au$^{-1}$. 
Together, the two constraints discussed above imply a strong surface density gradient from 30 au to 45 au. 
For example, if the linear mass density of the planetesimal disk followed $\exp[-(r-24\,{\rm au})/\Delta r]$, 
then $\Delta r \sim 2.5$ au. This is consistent with the total mass and radial profile of the planetesimal 
disk that we used for Figure \ref{case40}. 

\begin{figure}
\epsscale{0.6}
\plotone{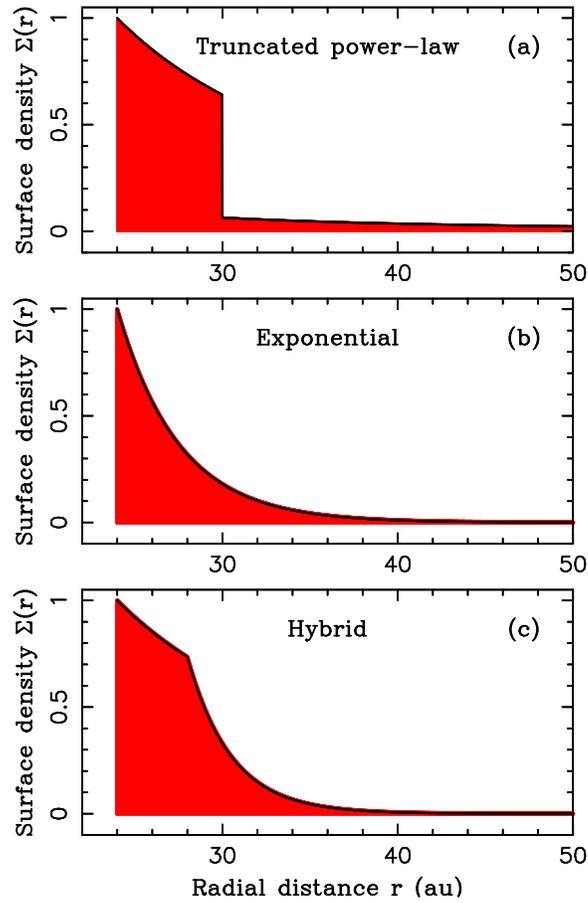}
\caption{Three planetesimal disk profiles used in this work: (a) truncated power law, (b) exponential, and (c) 
power-law inner disk and exponential outer disk. The surface density is arbitrarily normalized here to 
$\Sigma=1$ at 24 au.}
\label{weights}
\end{figure}

Here we examine three disk profiles: the (1) truncated power-law (surface density $\Sigma \propto 1/r^\gamma$ with 
$\gamma=1$-2, truncated at 30 au, a low-mass extension beyond 30 au; Fig. \ref{weights}a), (2) exponential ($\Sigma \propto \exp [(r-r_0)/\Delta r]/r$, 
where $r_0$ is the inner edge radius near $\sim$24 au, and $\Delta r$ is one e-fold, no outer 
truncation here; Fig. \ref{weights}b), and (3) hybrid profiles (power law $\Sigma \propto 1/r^\gamma$
below 28 au, exponential above 28 au; Fig. \ref{weights}c). Each of our simulations includes two 
million disk planetesimals distributed from outside Neptune's initial orbit at $\sim$24 au to 
$>$50 au (one million at $<30$ au and one million at $>30$ au). Such a large number of bodies is needed to obtain good 
statistics. The initial eccentricities and inclinations of planetesimals are set according to the Rayleigh distribution with $\sigma_e=0.1$ and 
$\sigma_i=0.05$. The planetesimals are assumed to be massless, such that their gravity does not interfere 
with the code's migration/damping routines. 

We use {\it weights}\footnote{These weights have no physical meaning. Thay are simply a way of tracking where 
particles start/end.} to set up the three profiles in Fig. \ref{weights}. Specifically, the planetesimals 
starting at orbital radius $r$ are given weight $w(r)$, where $w(r)$ follows the selected initial 
density profile. The weights are propagated through the simulation and analysis, and used to gauge the
contribution of each particle to the model results, including the color ratios reported in Sect. 4.3.
For the truncated power-law profile in Fig. \ref{weights}a, the step in the surface density at 30 au 
is parametrized by the contrast parameter, $c$, which is simply the ratio of densities on either side 
of 30 au. The exponential and hybrid disks in Figs. \ref{weights}b and c are parametrized by one e-fold 
$\Delta r$. To match the constraints described above, $\Delta r$ of the hybrid disk must be smaller 
than $\Delta r$ of the exponential disk. We assign a color to each simulated object depending on 
whether it started at $r<r^*$ or $r>r^*$. The color transition at $r^*$ is assumed to be a sharp boundary 
between R and VR. 

\subsection{Comparison with observations}

We use the OSSOS detection simulator (Bannister et al. 2018) to 
show that our model results are roughly consistent with the orbital structure of the 
Kuiper belt. A more systematic comparison will be published elsewhere as part of the OSSOS publication 
series. OSSOS is the largest Kuiper belt survey with published characterization (1142 ensemble detections; 
Bannister et al. 2018). The simulator was developed by the OSSOS team to aid the interpretation of their 
observations. Given intrinsic orbital and magnitude distributions, the OSSOS simulator returns a sample 
of objects that would have been detected by the survey, accounting for flux biases, pointing history, 
rate cuts and object leakage (Lawler et al. 2018). 

In this work, we input our model results into the OSSOS simulator to compute the detection statistics. We 
first increase the statistics by performing a 10-Myr integration starting from $t=4.5$ Gyr. The orbital 
elements of planets and model KBOs are saved with the $10^5$ yr cadence (generating 100 outputs for each body). For 
each output, we rotate the reference system such that Neptune appears near Neptune's current position on the sky. 
This assures a consistency with the OSSOS observations. The OSSOS simulator then reads the orbital elements 
of model KBOs on input. 

The input magnitudes/sizes of KBOs are modeled as broken power-law distributions (e.g., Fraser et al. 2014). 
If the original massive disk is really the main source of all dynamically hot populations in 
the Kuiper belt, their size distributions should be similar and should reflect the size distribution of 
the disk at the time of its dispersal by Neptune. This is because the collisional evolution in 
the present-day Kuiper belt is modest and has not altered the size distribution in the size range of KBOs 
detected by OSSOS (e.g., Nesvorn\'y \& Vokrouhlick\'y 2019, their Fig. 12). In more detail, the size 
distribution can be informed from Jupiter Trojans because Jupiter Trojans were presumably captured at 5.2 au 
from the same source (e.g., Morbidelli et al. 2005, Nesvorn\'y et al. 2013; but see Pirani et al. 2019), 
their size distribution has not evolved after capture (Nesvorn\'y et al. 2018) and is well characterized 
from observations down to diameters $D \simeq 5$ km (Wong \& Brown 2015, Yoshida \& Terai 2017).   

Specifically, we use the cumulative size distribution $N(>\!\!D) = N_{\rm break} (D/D_{\rm break})^{-q_{\rm big}}$ 
for $D>D_{\rm break}$, where $D_{\rm break}$ is the location of the break, $N_{\rm break}$
is the number of bodies with $D>D_{\rm break}$, and $N(>\!\!D) = N_{\rm break} (D/D_{\rm break})^{-q_{\rm small}}$ for 
$D<D_{\rm break}$. From observations of Jupiter Trojans and KBOs we set $D_{\rm break}\simeq100$ km, $q_{\rm small} 
\simeq 2$ and $q_{\rm big} \simeq 5$. We normalize $N_{\rm break}$ such that the whole planetesimal disk 
corresponds to 15-20 $M_{\rm Earth}$. This gives $N_{\rm break} \sim 6\times10^7$ bodies with a factor of 
$\sim$2 uncertainty (mainly due to the uncertain bulk density of Jupiter Trojans and KBOs). The size 
distribution of planetesimals beyond 40 au is discussed in Appendix A. For simplicity, we use the same 
albedo $p_{\rm V}=0.1$ for all populations to convert between size and absolute magnitude.

Once we make sure that the simulation results are roughly consistent with the number and orbital 
distribution of KBOs (Sections 4.1 and 4.2), we proceed by comparing them with the color data (Section 4.3). 
For that, we use the color database of Peixinho et al. (2015) (publicly available at\\ http://vizier.cfa.harvard.edu/), 
the color survey results of Wong \& Brown (2017) (356 objects, data available from authors), published data 
from the Col-OSSOS survey (Pike et al. 2017, Schwamb et al. 2019, Marsset et al. 2019), and the Neptune 
Trojan data from Jewitt (2018) and Lin et al. (2019). 

The results for each disk profile and $r^*$ are compared with the observations discussed above. 
Specifically, we require that: (1) at least $\sim$90\% of Neptune Trojans end up to be R 
(the best estimate of the intrinsic VR/R ratio of Neptune Trojans, $\sim$0.06 from Lin et al. 2019, has 
a large statistical uncertainty), (2) over $\sim$90\% of CCs end up to be VR (to reflect the 
predominantly VR colors among CCs), and (3) the VR/R ratio of dynamically hot KBOs (HCs, Plutinos, SDOs) 
obtained in the model matches the VR/R ratio inferred from observations (intrinsic VR/R $\sim 0.1$-0.3; 
Wong \& Brown 2017, Schwamb et al. 2019). The color distribution is reported here for $D>100$ km objects 
(absolute magnitudes $H<8.1$ for $p_{\rm V}=0.1$). Changes of the color ratio with size are not investigated 
here. Special attention is given to the correlation of colors with orbital inclination (e.g., Peixinho et al. 2008, 
Marsset et al. 2019), which is the chief supporting argument for our color hypothesis (Section 4.3).

\subsection{Model caveats}

Here we adopt a numerical model with disk planetesimals that do not carry any mass. Neptune's migration 
in the planetesimal disk is mimicked by artificial forces. This is not ideal for several 
reasons. For example, this means that the precession frequencies of planets are not affected by the 
disk torques in our simulations, while in reality they were (Batygin et al. 2011). The direct gravitational 
effect of the fifth giant planet (NM12) on the disk planetesimals is also ignored. These additional effects,
which may influence the orbital structure of the Kuiper belt, are studied elsewhere (Nesvorn\'y et al., 
in preparation). 

The weighting scheme described in Sect. 3.2 is used to investigate the effect of different 
radial profiles of the planetesimal disk on the orbital and color distributions of KBOs. This scheme 
can strictly be applied only to the outer part of the disk beyond 30 au. For $r<30$~au, the 
radial disk profile is tied to Neptune's migration, and changing the profile would mean that the 
character of Neptune's migration would change as well. Given the high computational expense of 
these calculations, however, we are unable to resolve this dependence in full detail. So, we do 
the next best thing, which is to use the weighting scheme to capture the dependence on the source 
material initially available at different orbital radii, and different migration cases to capture the 
dependence on the nature of Neptune's migration. The development of a more self-consistent model 
that would account for coupling between these effects is left for future work.  
 
Ideally, we would like to have a large suite of dynamical models with different migration timescales, 
different histories of Neptune's orbit (Volk \& Malhotra 2019), etc., and choose the best model by 
formally fitting the OSSOS dataset. 
This would also help to establish the uncertainty of model parameters. Such a systematic exploration 
of model parameters, however, is not possible with only four dynamical models available to us (Table 1). 
Here we therefore focus on establishing trends with different parameters in an attempt to roughly 
triangulate the interesting range of possibilities. The main goal of these efforts is to demonstrate 
the plausibility of the color hypothesis proposed in Sect. 2. 

\section{Results} 

\begin{figure}
\epsscale{0.8}
\plotone{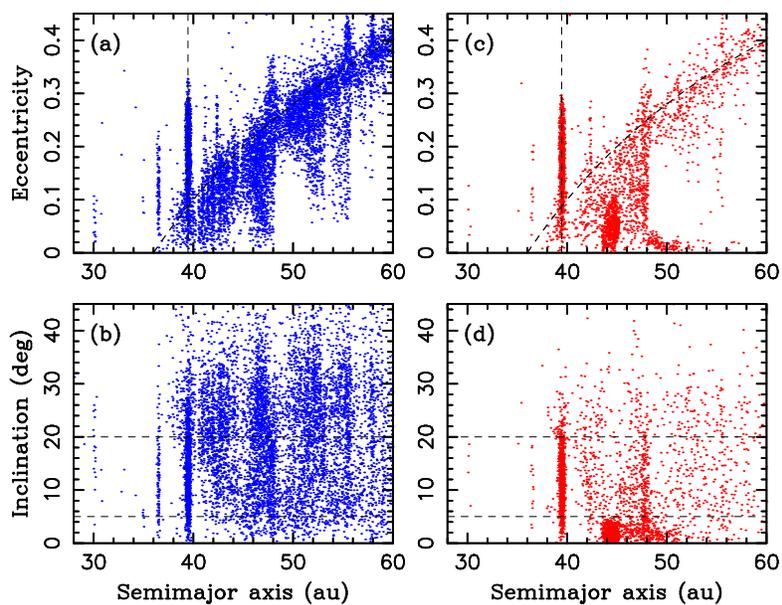}
\caption{The raw orbital distribution of KBOs obtained in the s30/100j model. The final orbits of 
planetesimals starting with $a<35$ au ($a>35$ au) are plotted as blue (red) dots in panels a and b (c and d). 
The dashed lines show the location of the 3:2 resonance with Neptune ($a=39.45$ au in a and c), perihelion 
distance $q=36$ au (a and c), and inclinations $i=5^\circ$ and 20$^\circ$ (panels b and d).}
\label{raw1}
\end{figure}

The raw orbital distribution of bodies implanted into the Kuiper belt ($t=4.5$ Gyr) is shown in Figure 
\ref{raw1}. Note that the raw distribution should not be directly compared to observations because: it (1) 
does not account for observational biases and (2) corresponds to disk profile with $w(r)=1$ 
for all $r$ (which is unphysical; e.g., too much emphasis is given to bodies starting with $r>30$ au).
We show these plots to illustrate a typical result of our simulations. There are several notable 
features. The resonant populations, including Neptune Trojans, Plutinos and objects in the 4:3, 2:1 and 5:2 
resonances can clearly be identified. Interestingly, the planetesimals sourced from $r<35$ au tend to evolve 
to higher orbital inclinations than the ones from $r>35$ au (compare panels (b) and (d) in Fig. \ref{raw1}). 
We interpret this trend as a consequence of weakening of Neptune's gravitational perturbations with $r$ and use it to discuss 
the color-inclination correlation in Sect. 4.3. 

Another prominent feature in Fig. \ref{raw1}c,d is the concentration of bodies with $a\simeq44$ au, 
$e<0.1$ and $i<5^\circ$. The concentration appears in the simulation when planetesimals starting 
at $\sim$40-43 au are captured into the 2:1 resonance with Neptune and subsequently released from the resonance 
during Neptune's jump. The slow migration of Neptune's 2:1 resonance after the jump depletes the cold population 
beyond 45 au. Results similar to these were used in Nesvorn\'y (2015b) to explain the Kuiper belt kernel
(Petit et al. 2011). We will discuss the kernel in more detail in Sect. 4.2. Here we just note that models 
with a continuous migration of Neptune (i.e., no jump) produce a dispersed orbital distribution of CCs 
and no kernel. 

\subsection{Number of KBOs}

The number of bodies in different Kuiper belt populations has been determined from observations. 
Petit et al. (2011) estimated that there should be $35,000\pm8,000$ HCs with $D>100$ km. Gladman et al. 
(2012) found that the population of Plutinos in the 3:2 resonance represents $\sim$1/3 of the HC population. 
From this we have $\sim$12,000 Plutinos with $D>100$ km. Petit et al. (2011) also estimated that there should 
be $\sim$95,000 CCs with $D>100$ km, but such a large population would be at odds with Fraser et al. (2014), 
who found that the total mass of CCs represents only $\sim$0.03 of the HC population mass. We find 
$\simeq$15,000 CCs with $D>100$ km from OSSOS (Appendix A). Finally, Lin et al. (2018) estimated from 
the Deep Ecliptic Survey that there should be $\simeq$160 Neptune Trojans with absolute magnitude $H<10$ 
($D>50$ km for albedo $p_{\rm V}=0.07$) in the $L_4$ point. Assuming that the $L_5$ population is similar and 
approximately re-scaling to $D>100$ km, we find that there should be $\sim$100 Neptune Trojans with $D>100$ 
km. If so, the population of Neptune Trojans (NTs) would be roughly four times larger than the population 
of Jupiter Trojans.

We now address the question of how well the migration models investigated in this work reproduce the 
inferred number of objects in different populations. The s30/100j simulation yields reasonable results 
(Fig. \ref{pops}). Some of the best results are obtained in this case with the exponential profile 
and $\Delta r = 2.5$ au, where we identify 40,000 HCs, 25,000 Plutinos, 15,000 CCs and 300 NTs with 
$D>100$ km (here we assume that the original disk contained $6\times10^7$ $D>100$ km planetesimals; 
Nesvorn\'y 2018).\footnote{The model also gives 10,000 $D>100$ km bodies in the 2:1 resonance and 
150,000 $D>100$ km SDOs. See, for example, Nesvorn\'y (2018) for a discussion of these populations.}
The s30/100j simulation with the step profile works as well. Specifically, with $c=1000$, 
we obtain 35,000 HCs, 20,000 Plutinos, 20,000 CCs and 200 NTs with $D>100$ km, which is practically the 
same result as for the best exponential profile discussed above. The hybrid disk profiles yield intermediate 
results and we do not highlight them here.

The s30/100 model (no jump) does not work with the exponential and hybrid profiles. This is 
because in both these cases, $\Delta r > 2$ au is needed to obtain a roughly correct number of CCs. 
But $\Delta r > 2$ au also gives a very large number of Plutinos, both in the absolute ($>$70,000 bodies 
with $D>100$ km) and relative terms (the Plutino population ends up to be larger than the HC population). 
This is a consequence of the 3:2 resonance sweeping and efficient capture of planetesimals from 
$r \simeq 32$-39 au (Hahn \& Malhotra 2005). The truncated power-law profile can be used to resolve this problem. 
Indeed, with $c=1000$, we obtain 40,000 HCs, 25,000 Plutinos, 15,000 CCs and 450 NTs with $D>100$. This 
is identical to the s30/100j case with $\Delta r = 2.5$ au, except that there are $\sim$1.5 times more 
NTs in s30/100. We do not consider this to be a problem because the population of NTs is very sensitive to 
details of Neptune's migration history (e.g., Gomes \& Nesvorn\'y 2016).  

The s10/30 (no jump) model gives results that are much less compatible with the structure of 
the Kuiper belt than the results discussed above. This applies independently of the assumed radial profile.
For example, with the truncated power-law profile and any value of $c$, the s10/30 model gives an 
excessive number of HCs (over 80,000 with $D>100$ km), Plutinos (over 60,000) and NTs (over 2,000). We  
do not include the s10/30 model in the following analysis. The results are better when the jump is applied
to Neptune's orbit in the s10/30j model, because the jump acts to lower the population of NTs and 
Plutinos. For example, with the step profile and $c=1000$, we obtain
70,000 HCs, 40,000 Plutinos, 20,000 CCs and 1,000 NTs with $D>100$ km.  These population estimates 
are at least a factor of $\sim$2 higher than what we inferred from the observational constraints above. 
The s10/30j model would potentially be plausible if the original disk contained fewer than $6\times10^7$ 
$D>100$~km planetesimals. The results for s10/30j and the exponential profile with $\Delta r=2.5$ au
are similar.

In summary, we identified the following cases that reasonably well reproduce the number of KBOs in 
different populations: s10/30j and s30/100j with the exponential profile and $\Delta r \simeq 2.5$ au, and
s10/30j, s30/100j and s30/100 with the truncated power-law profile and $c \simeq 1000$. The hybrid 
profiles are plausible as well. 
 
\subsection{Orbital distribution}

Here we examine the orbital distribution of KBOs. Our goal is to 
show that the dynamical models reproduce the observed structure reasonably well and can thus be 
used to investigate Kuiper belt colors. A more detailed statistical analysis of the orbital 
distribution will be published elsewhere. 

\begin{figure}
\epsscale{0.5}
\plotone{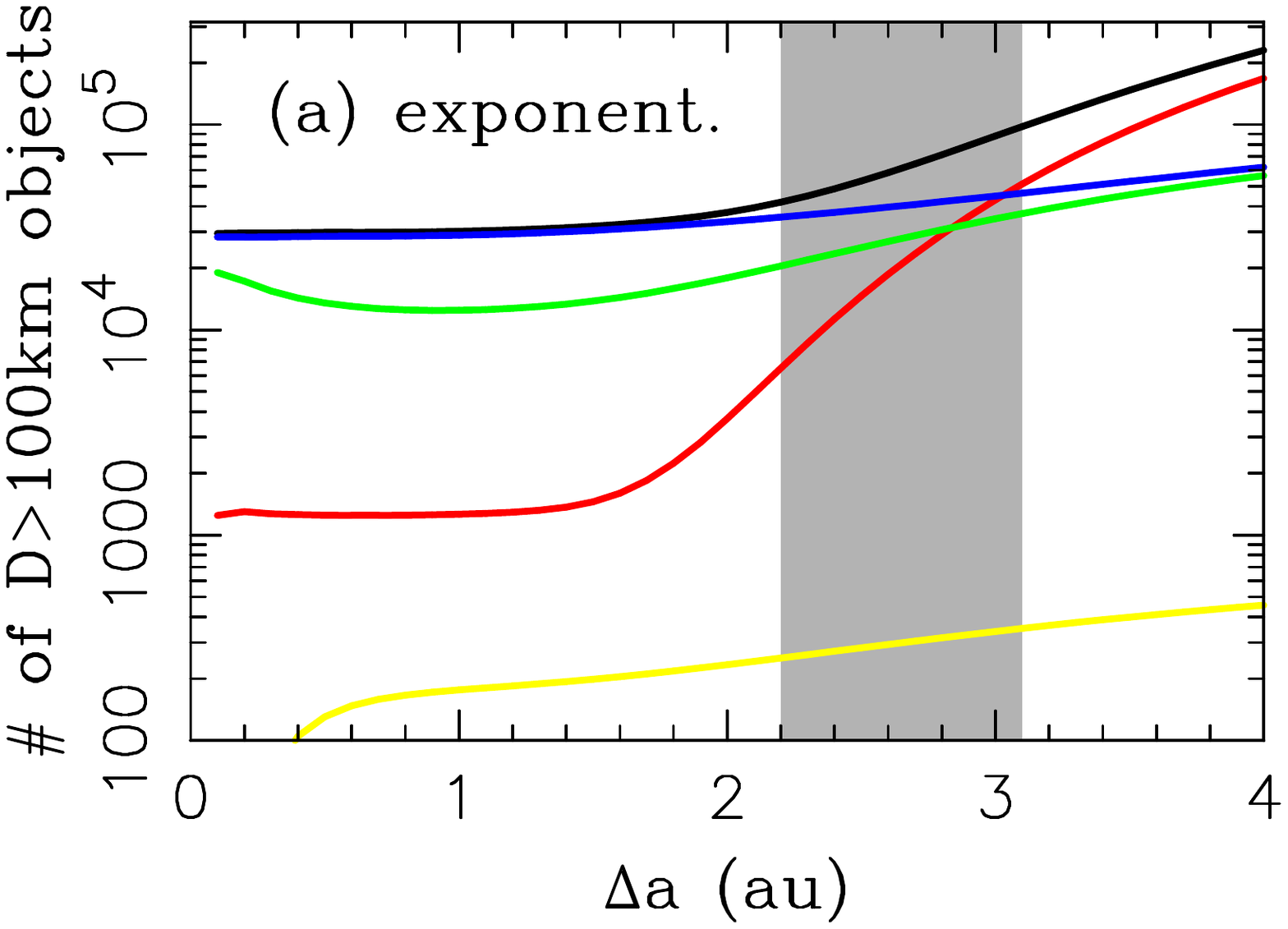}%\\[2.mm]
%\hspace{1.mm}\epsscale{0.32}
\plotone{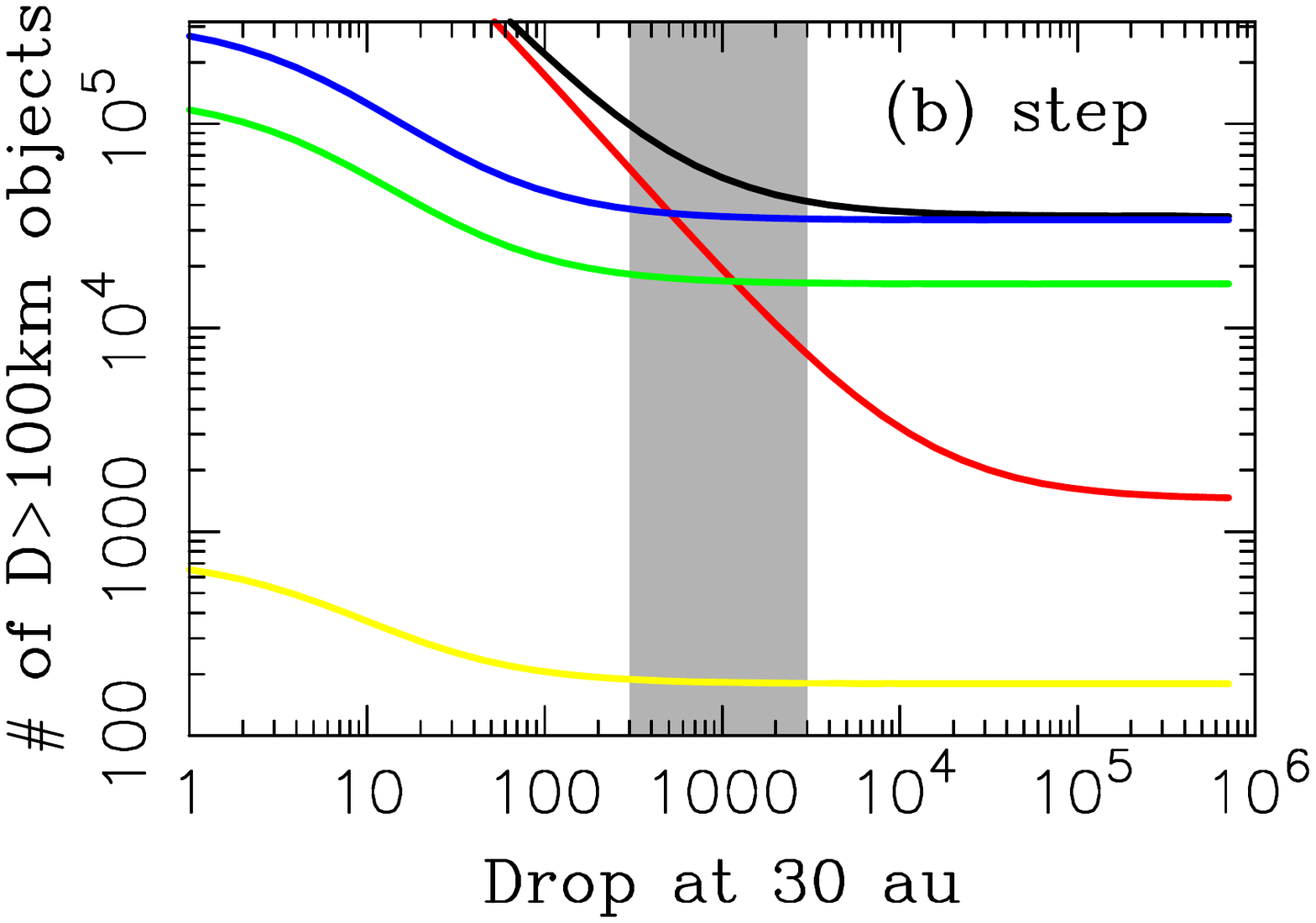}
\caption{The population estimates from the s30/100j model for different initial profiles of the original 
planetesimal disk. Different lines correspond to the main CKB (black), HCs (blue), CCs (red), 
Plutinos (green) and Neptune Trojans (yellow). The top panel is the exponential profile starting at $r_0=24$ 
au. The bottom panel is the truncated power-law profile with a surface density drop at 30 au. The shaded areas 
highlight the plausible range of parameters. The exponential profiles with $\Delta r<2.2$ au can be ruled out 
because the population of CCs predicted in such models is too small ($<$7,000 $D>100$ km bodies; most CCs would be 
deposited onto their present orbits from $r<30$ au if $\Delta r<1.5$ au). Similarly, the exponential models with 
$\Delta r>3.1$ au can be ruled because the predicted number of CCs is excessive ($>$50,000 
$D>100$ km bodies for $\Delta r>3.1$ au). Similar arguments based on the CC population can be used to rule 
out $c<300$ and $c>3000$ for the case shown in panel~(b).}
\label{pops}
\end{figure}

\begin{figure}
\epsscale{0.49}
\plotone{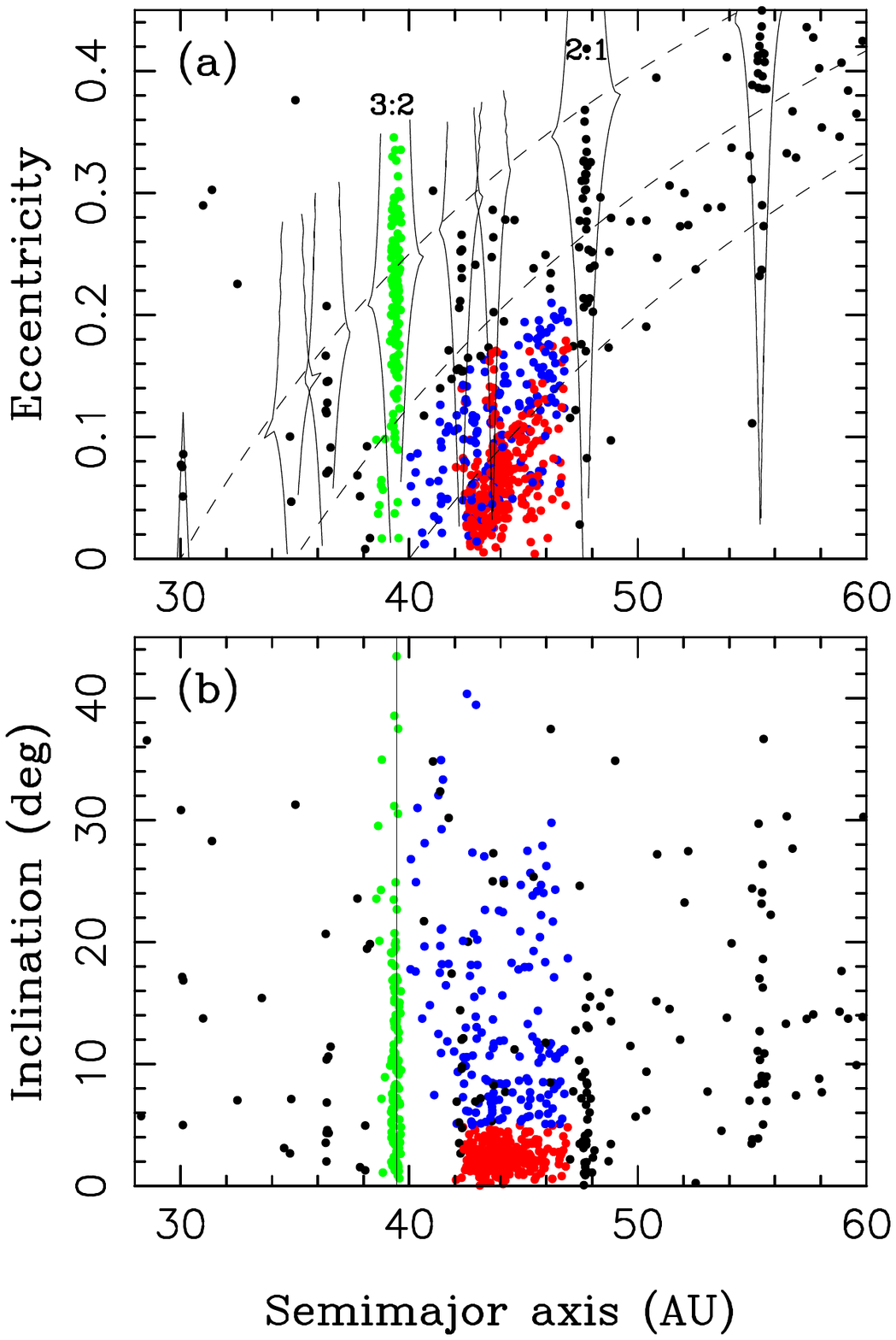}
\plotone{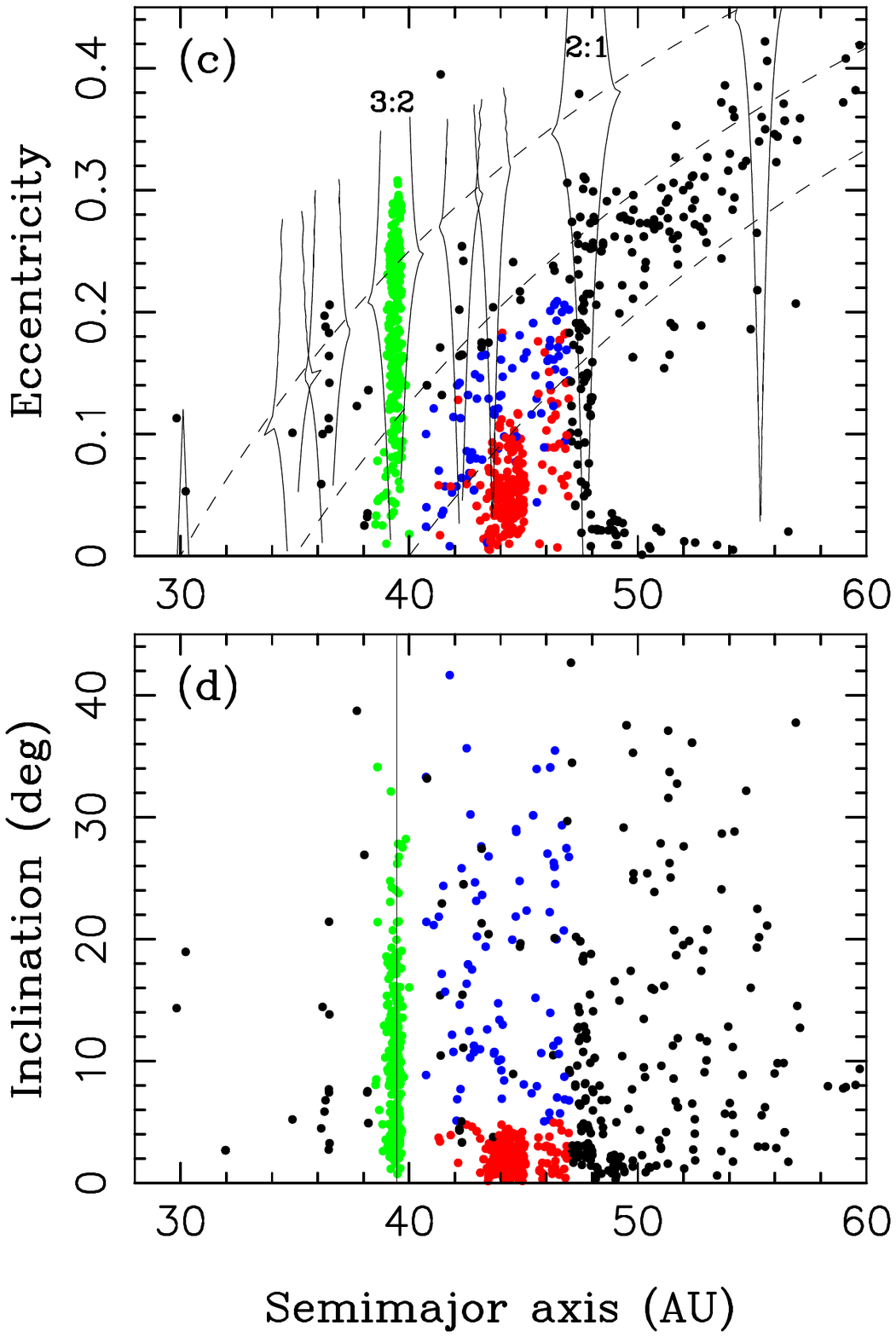}
\caption{The OSSOS KBOs (left panels) and tracked detections from the s30/100j model with $\Delta r=2.5$~au
(right panels). In both cases, the plots show the barycentric orbital elements referred to the center of mass 
of the Solar System. For ease of comparison different populations are highlighted by different colors:
Plutinos (green dots), HCs (blue dots) and CCs (red dots). The orbital elements of all other KBOs are shown
as black dots.}
\label{track}
\end{figure}

Figure \ref{track} compares the OSSOS detections of KBOs (844 objects in total; some fall outside the 
plotted range) and tracked detections from the s30/100j model. To generate the model distribution in 
Fig. \ref{track}c,d, we use the raw orbital distribution from Fig. \ref{raw1} and weights 
corresponding to the exponential disk profile with $r_0=24$ au and $\Delta r = 2.5$ au. This profile 
gives the correct number of objects in different KBO populations as discussed in the previous section 
(Fig. \ref{pops}a). To simulate the OSSOS detections, we adopt $D_{\rm break}=100$ km, $q_{\rm small}=2.1$
$q_{\rm big}=5$ and $p_{\rm V}=0.1$, and instruct the OSSOS simulator to detect and track 844 objects 
in total (the same as the total number of OSSOS detections).

The s30/100j model works well to reproduce the general orbital structure of the Kuiper belt. 
The model distribution shows populations of bodies in all main orbital resonances with Neptune, 
including Neptune Trojans in 1:1, Plutinos in 3:2 and Twotinos in 2:1. There are also tracked 
detections in the inner 4:3 resonance, the CKB resonances (e.g., 5:3, 7:4, 8:5), and the outer 5:2
resonance. An interesting difference between the OSSOS and model distributions is noted for the
2:1 and 5:2 resonances, where OSSOS detected important ``core'' populations (Fig. \ref{track}a,b;
Volk et al. 2016, Chen et al. 2019).
These populations are present in the raw results (Fig. \ref{raw1}) but are not sufficiently 
large to stand out in model detections (Fig. \ref{track}c,d). This issue is not related
to Neptune's jump because the same problem exists without a jump in the s30/100 simulation.
The model with faster Neptune's migration, s10/30j, performs better in this respect in that it 
shows the core populations in the 2:1 and 5:2 resonances, which are only factor of $\sim$2 
below the number of objects actually detected by OSSOS. 

The classical belt produced in the s30/100j model closely replicates the OSSOS CKB detections.
There are HCs with a broad inclination distribution and CCs with low inclinations. The Kuiper
belt kernel is notable in Fig. \ref{track}c,d and its orbital structure is very similar to 
that shown in Fig. \ref{track}a,b. We compare the orbital distributions of CCs in more detail below.
There is a tail of low-eccentricity and low-inclination orbits beyond the 2:1 resonance which 
does not have a counterpart in the OSSOS detections. This may indicate that the radial profile 
of the original disk was steeper at 45-50 au than the one used here with $\Delta r = 2.5$ au. 
For example, if we use $\Delta r = 2$ au instead, the tail population near 50 au is reduced 
by a factor of $\sim$5, which would be more in line with OSSOS observations. This would 
also imply a much smaller CC population (Fig. \ref{pops}a). 

\begin{figure}
\epsscale{0.55}
\plotone{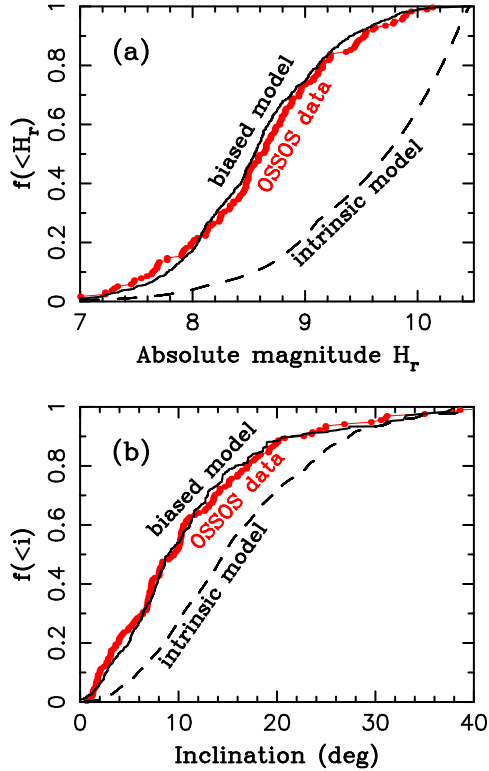}
\caption{A comparison between the biased model (solid black line) and OSSOS observations (red dots) 
of Plutinos in the 3:2 resonance with Neptune. Following the method described in Section 3 we simulate 
resonant capture of Plutinos from the original planetesimal disk. The OSSOS simulator was then applied 
to the present-day intrinsic model distribution (dashed lines). The large difference between the 
intrinsic and biased distributions illustrates the extreme biases of KBO observations and the importance 
of the OSSOS simulator. These results were obtained with an exponential disk profile ($r_0=24$ au, 
$\Delta r = 2.5$ au, $M_{\rm disk}=20$ $M_{\rm Earth}$). See Fig. \ref{case40} for Neptune's migration 
in this case. The (biased model) absolute magnitude (panel a) and orbital inclination (panel b) 
distributions match OSSOS observations. The results for orbital eccentricities and libration amplitudes, 
not shown here, are equally good.}
\label{plutino}
\end{figure}

Figure \ref{plutino} compares the orbital distributions for Plutinos. The intrinsic inclination 
distribution of Plutinos is very broad with the median $\simeq$15$^\circ$. It matches, after being biased 
by the OSSOS simulator, the OSSOS detections. Here we used $\Delta r = 2.5$ au. The s30/100j 
model with a truncated power-law profile and $c\simeq1000$ works equally well. This means that 
the radial profile of the original planetesimal disk cannot be inferred from the orbital 
distribution of Plutinos alone. This information was lost during the implantation process. 
The s10/30j models with $\Delta r = 2.5$ au or $c\simeq1000$ produce similar results, but
the biased model inclination distribution is slightly narrower than the observed one. 
A general correlation between the inclination distribution of KBOs and Neptune's migration 
timescale was pointed out in Nesvorn\'y (2015a; but see Volk \& Malhotra 2019).

The OSSOS inclination distribution of HCs shows a potential break near 12$^\circ$.
Below this break the distribution is steeply raising such that approximately 
60\% of detected HCs have orbital inclinations below the break. The distribution is shallower 
above the break and extends to $i>30^\circ$. This feature has previously been noted in Nesvorn\'y 
(2015a) and interpreted as a consequence of Neptune's migration into an extended disk. Here 
we find that the high-$i$ part of the distribution is implanted from $r\lesssim35$ au
(implying more excitation) and the low-$i$ part started at $r\gtrsim35$ au (implying
less excitation). The inclination distribution of HCs may therefore help to constrain the 
radial profile of the original disk. The best results are obtained with $\Delta r > 3$~au or $c < 500$
for the s30/100j model creating some tension with our general preference for $\Delta r \simeq 
2.5$ au or $c \simeq 1000$. The s10/30j models with $\Delta r \simeq 2.5$ au and 
$c \simeq 1000$ produce narrower inclination distributions. This may indicate 
that the actual migration timescale of Neptune was intermediate between the two timescales 
investigated here. 

\begin{figure}
\epsscale{0.55}
\plotone{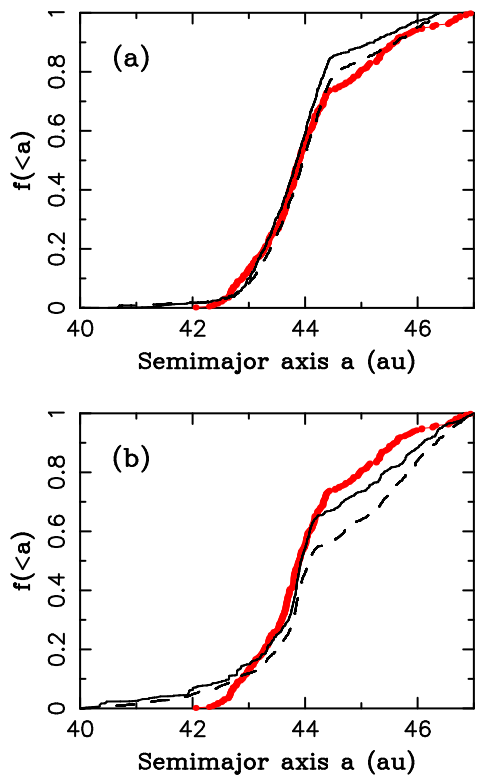}
\caption{A comparison between the biased models (solid black lines) and OSSOS observations (red dots) 
of CCs ($i<5^\circ$). The intrinsic semimajor axis distributions obtained in our models are shown by 
dashed lines. Panel (a) shows the result for the s10/30j model and panel (b) shows the result for the s30/100j
model. The Kuiper belt kernel is represented by a steeply raising semimajor axis distribution 
toward a break near 44.3 au.}
\label{kernel}
\end{figure}

The s10/30j and s30/100j models produce the Kuiper belt kernel (Petit et al. 2011, Nesvorn\'y 2015b). 
The kernel appears in the OSSOS observations as a strong concentration of low-$i$
KBOs near 44 au (Fig. \ref{kernel}; also see Fig. \ref{track}). The observed kernel has an 
outer edge at $\simeq$44.3 au, beyond which the number density of CCs drops by a factor of $\sim$5.
The migration models with Neptune's jump work well to reproduce these observations. The s10/30j model leads to a slightly 
stronger concentration of bodies below $\simeq$44.3 au (Fig. \ref{kernel}a), whereas the 
s30/100j model leads to a slightly weaker concentration (Fig. \ref{kernel}b). Again, this may 
indicate that the actual evolution of Neptune's orbit was intermediate between our two models.

\subsection{Colors}

\begin{figure}
\epsscale{0.5}
\plotone{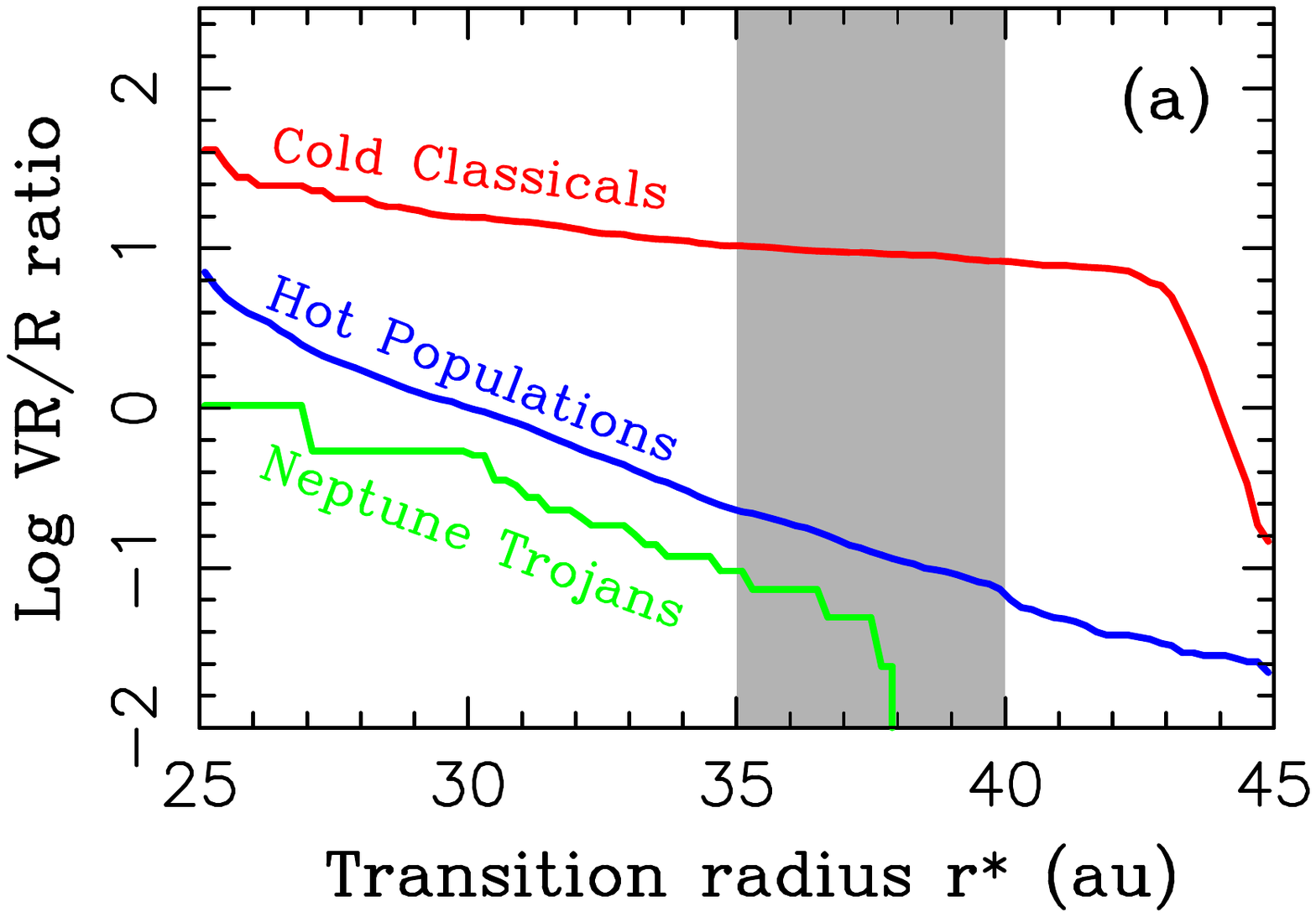}%\\[5.mm]
\plotone{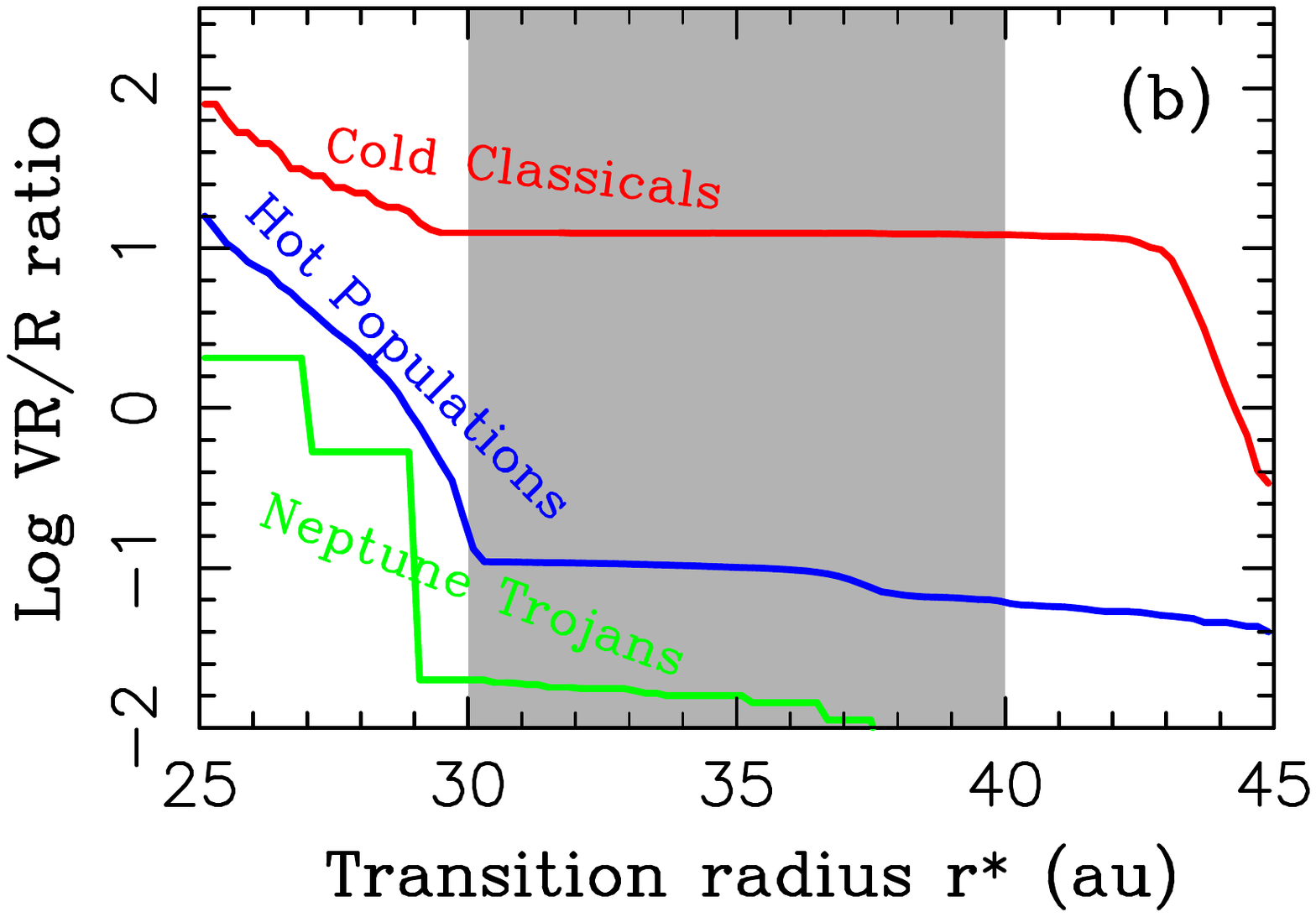}
%\epsscale{0.52}
%\hspace{1.mm*}\plotone{rstar_c1p3.eps}
\caption{The intrinsic VR/R ratio in the s30/100j model for different initial profiles of the original 
planetesimal disk: the (a) exponential profile with $\Delta r = 2.5$ au, and (b) truncated power-law 
profile with $c=1000$. Different lines correspond to CCs (red), hot populations (blue) and Neptune 
Trojans (green). The shaded area approximately highlights the possible range of VR/R transition radii in 
the original disk.}
\label{rstar}
\end{figure}

\begin{figure}
\epsscale{0.5}
\plotone{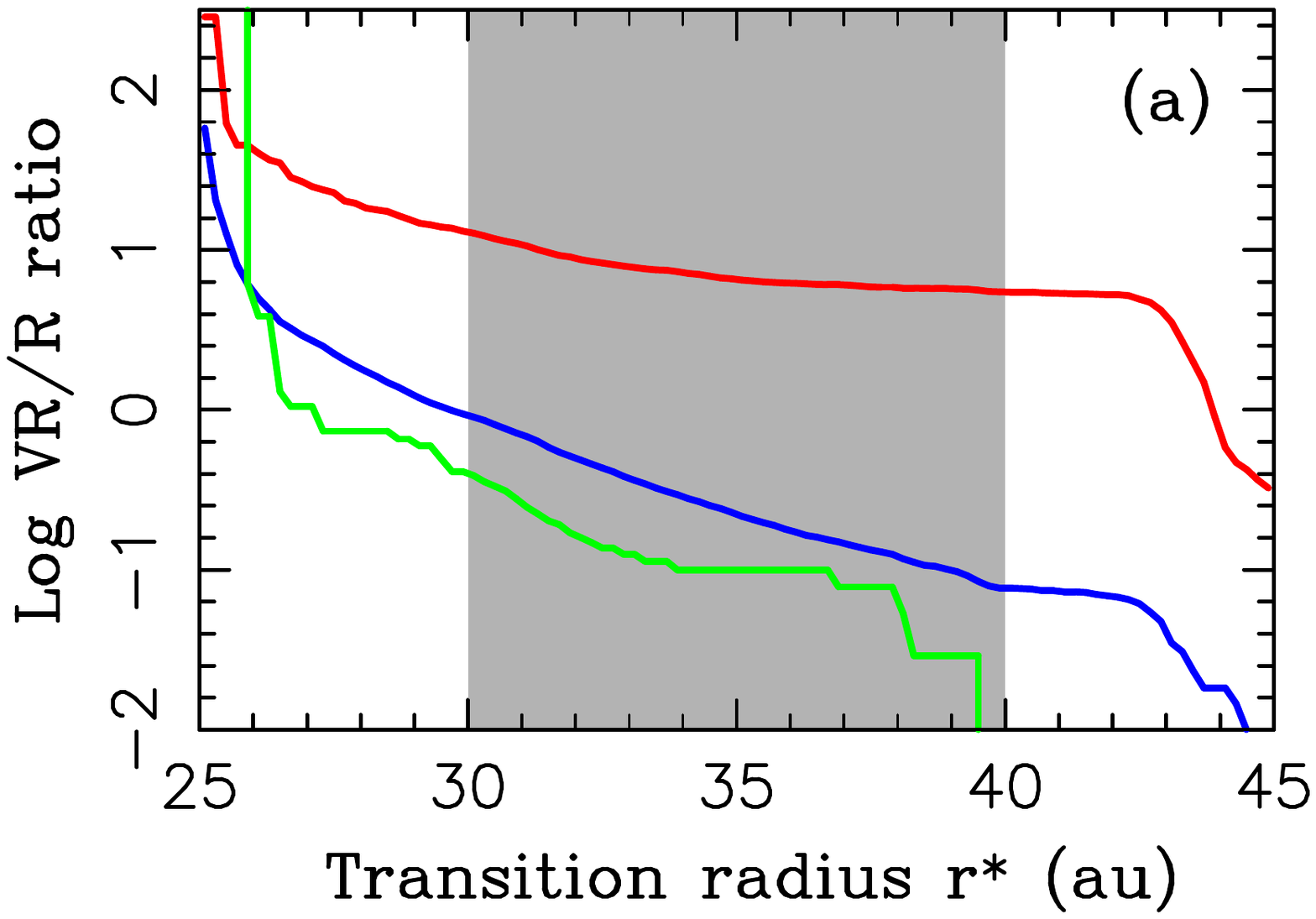}%\\[5.mm]
\plotone{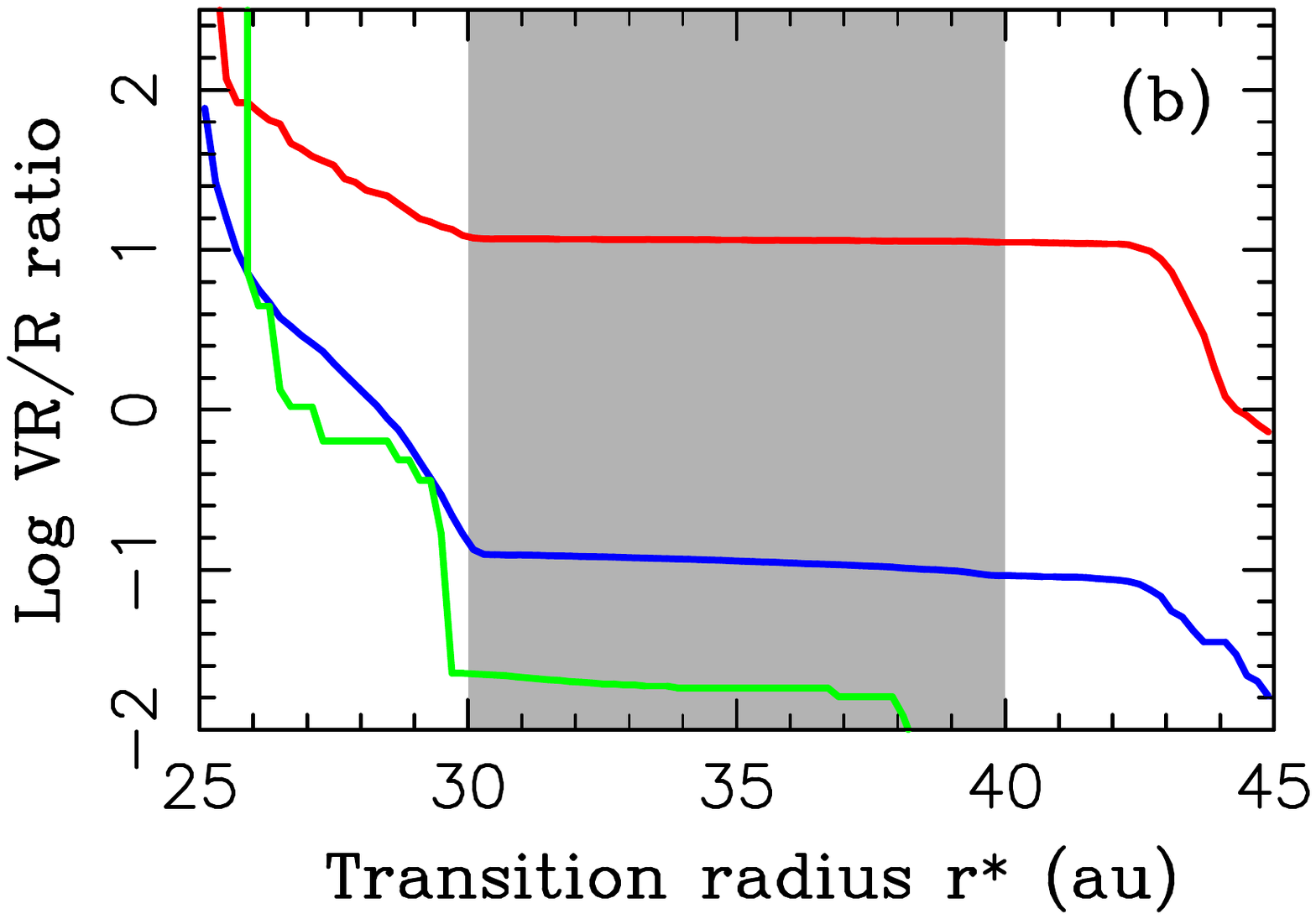}
%\epsscale{0.52}
%\hspace{1.mm*}\plotone{rstar_c1p3.eps}
\caption{The intrinsic VR/R ratio in the s10/30j model for different initial profiles of the original 
planetesimal disk: the (a) the exponential profile with $\Delta r = 3$ au, and (b) truncated power-law 
profile with $c=300$. The different lines correspond to CCs (red), hot populations (blue) and Neptune 
Trojans (green). The shaded area approximately highlights the possible range of VR/R transition radii in 
the original disk.}
\label{rstar2}
\end{figure}

The intrinsic VR/R color ratios in different KBO populations are plotted for the s30/100j and
s10/30j models in Figs. \ref{rstar} and \ref{rstar2}, respectively. In Fig. \ref{rstar}, we 
set $\Delta r = 2.5$ au (panel~a) and $c=1000$ (panel b) and plot the VR/R ratio as a function
of the transition radius $r^*$. As expected, the VR/R ratios are inversely correlated with $r^*$
(i.e., lower VR/R values are obtained for larger transition radii). The profiles for the truncated 
power-law disks show a slope change near 30 au, which is a reflection of the surface density 
discontinuity in these models. The exponential disk models lead to a more continuous change of
the VR/R ratio with the transition radius. In both cases, CCs show a nearly constant VR/R ratio
for $30<r^*<43$ au and, for obvious reasons, a sudden drop just outside of 43 au. There is not much 
difference between the s10/30j and s30/100j models. We do not have sufficiently good statistics for NTs, 
because the number of bodies captured as Neptune's co-orbitals is generally small. That is why the color 
ratios of NTs in Figs. \ref{rstar} and \ref{rstar2} are choppy.

Comparing these results with the color constraints discussed in Sect. 3.3 we find that $35<r^*<40$ au 
for the exponential profile in Fig. \ref{rstar}a and $30<r^*<40$ au for the truncated power-law profile in 
Fig. \ref{rstar}b (s30/100j model). For example, using the exponential profile and $r^*=37$ au we find 
${\rm VR/R} = 10$ for CCs\footnote{The color ratio is obtained here for the OSSOS-inferred CC population 
discussed in Appendix A. If, instead, we changed our model to approximate the much smaller 
CC population from Fraser et al. (2014), the model VR/R ratio of CCs would be much lower, because R objects 
implanted from $r<r^*$ to $a \sim 45$ au and $i<5^\circ$ would represent a greater share of CCs. This ``pollution'' 
problem does not have simple dynamical solution and indicates that the CC population may have been 
sub-estimated in Fraser et al. (2014).}, ${\rm VR/R} = 0.15$ for the hot populations, and ${\rm VR/R} = 0.05$ 
for NTs. The truncated power-law profile and $r^*=35$ au lead to ${\rm VR/R} = 10$ for CCs, ${\rm VR/R} = 0.1$ 
for the hot populations, and ${\rm VR/R} = 0.01$ for NTs. 
 
The general trends for the s10/30j model are similar (Fig. \ref{rstar2}). Here we opted for using 
the exponential profile with $\Delta r = 3$ au (panel a) and the truncated power-law profile with
$c=300$ (panel b), because these parameters better match the color constraints. Specifically,
for $\Delta r = 3$ au and $r^*=35$ au, we obtain ${\rm VR/R} = 7$ for CCs, ${\rm VR/R} = 0.3$ 
for hot populations, and ${\rm VR/R} = 0.06$ for NTs. If, $c=300$ and $r^*=35$ au
instead, we get ${\rm VR/R} = 10$ for CCs, ${\rm VR/R} = 0.1$ for hot populations, and 
${\rm VR/R} = 0.02$ for NTs. 
All these cases therefore satisfactorily replicate the predominance of the VR colors among CCs, 
and the predominance of the R colors in other KBO populations. The color transition at $r^*<30$ au can 
be ruled out because the VR/R ratio would be generally too high. The color transition at $r^*>40$ au 
is not supported as well because the VR/R ratio would be generally too low. In summary, we find
that $30<r^*<40$ au. 

\begin{figure}
\epsscale{0.4}
\plotone{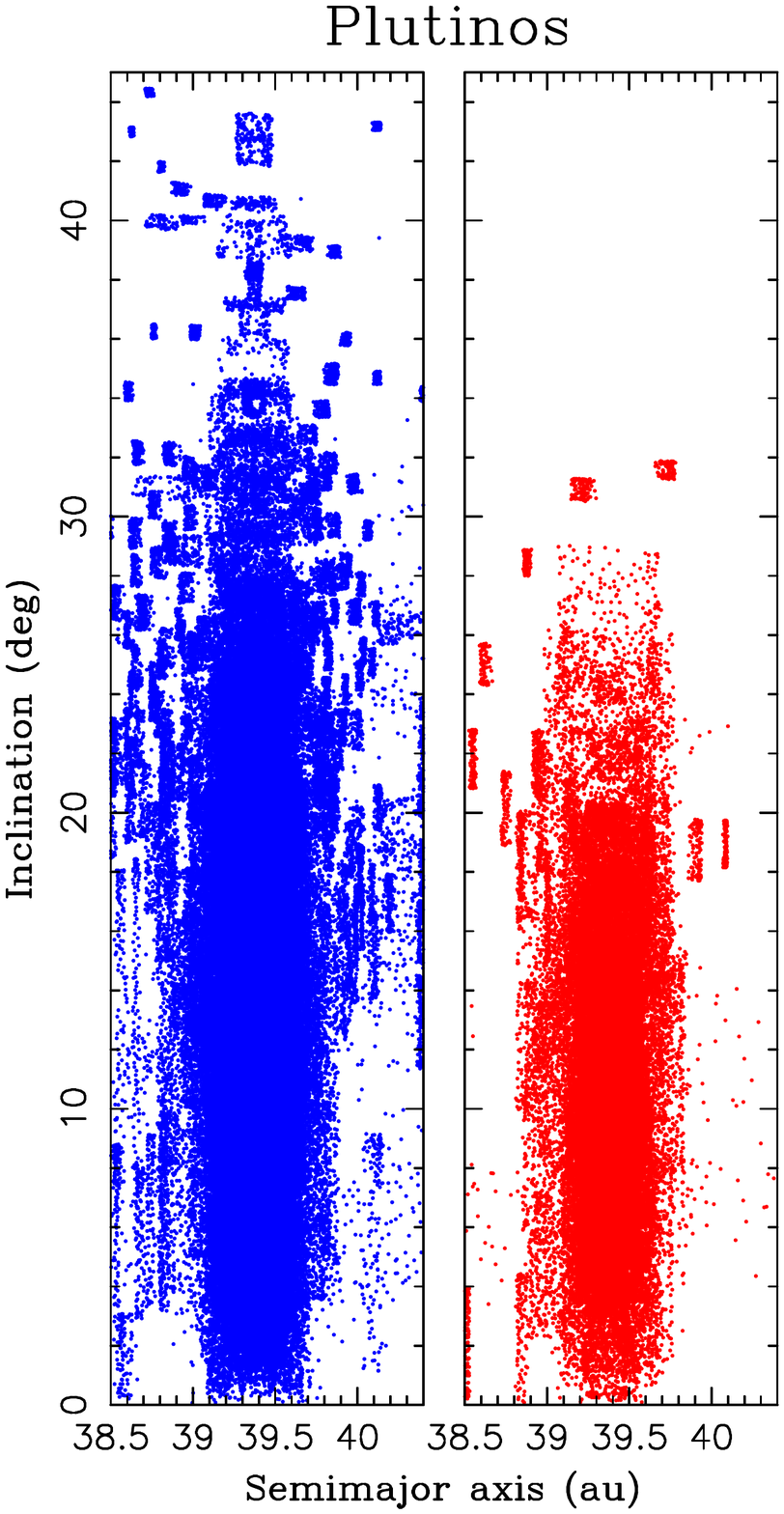}%\hspace{1.mm}
\plotone{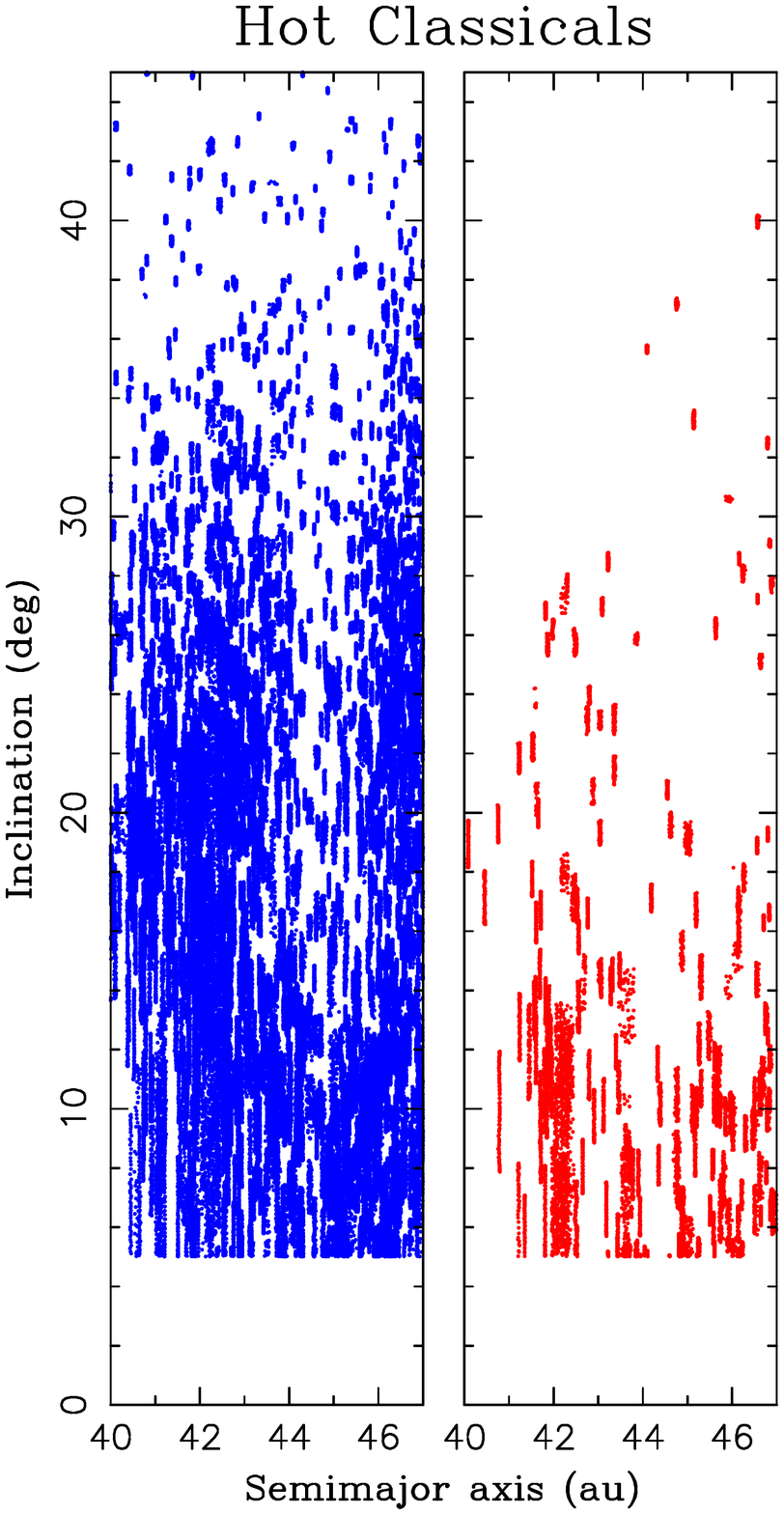}%\hspace{1.mm}
\plotone{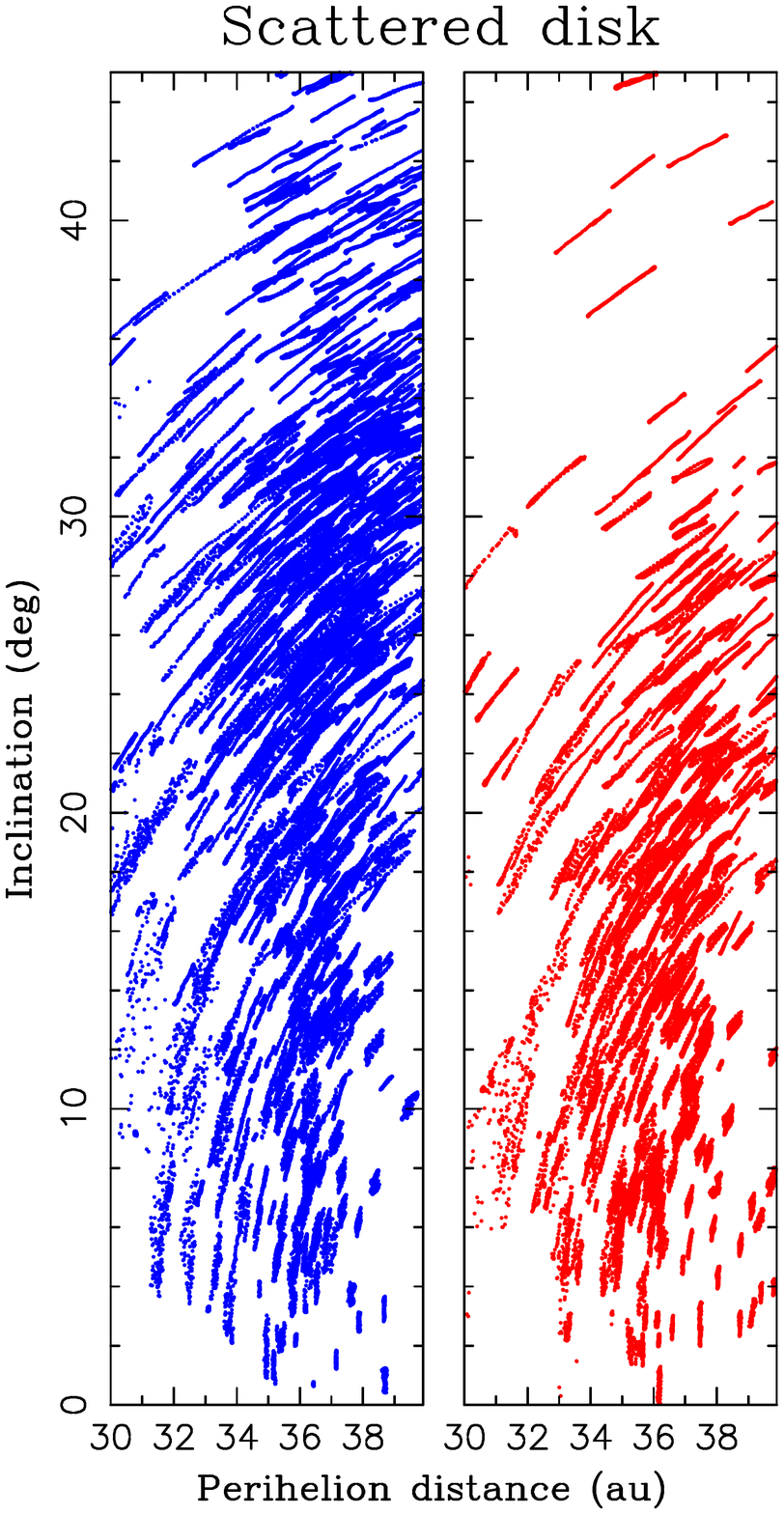}
%\epsscale{0.52}
%\hspace{1.mm*}\plotone{rstar_c1p3.eps}
\caption{The inclination distribution of R (blue dots) and VR (red dots) bodies obtained in the
s30/100j model with $\Delta r=2.5$ au and $r^*=37$ au. We show the distributions for Plutinos
(left panels), HCs (middle panels) and SDOs (right panels). For each individual object, we plot 
100 orbits from the 10 Myr integration starting at $t=4.5$ Gyr (see Section 3.3).
Only orbits with $i>5^\circ$ are plotted in the middle plots to avoid confusion with CCs. In all 
three cases, the inclination distribution of R objects is broader than the inclination distribution 
of the VR objects. Compare to Fig. 3 in Marsset et al. (2019).}
\label{marsset}
\end{figure}

We now consider the correlation between colors and orbital inclinations (e.g., Marsset et al. 2019).
For that, we use the s30/100j model with the exponential profile and $\Delta r = 2.5$~au, and
$r^*=37$ au (the results for other cases are similar, given that $30<r^*<40$ au). Bodies  
are separated into two color groups: R for bodies starting with $r<37$ au and VR for $r>37$~au. We plot the
final orbits in Fig. \ref{marsset}. The figure shows that the R objects in each population
--Plutinos, HCs and SDOs-- are expected to have wider inclination distribution than the VR objects. The R objects
starting in the inner disk have experienced, on average, stronger orbital perturbations
from Neptune and ended with higher inclinations. This explains the observed correlation between 
colors and orbital inclinations, and provides strong support for the color hypothesis proposed here.

\begin{figure}
\epsscale{0.4}
\plotone{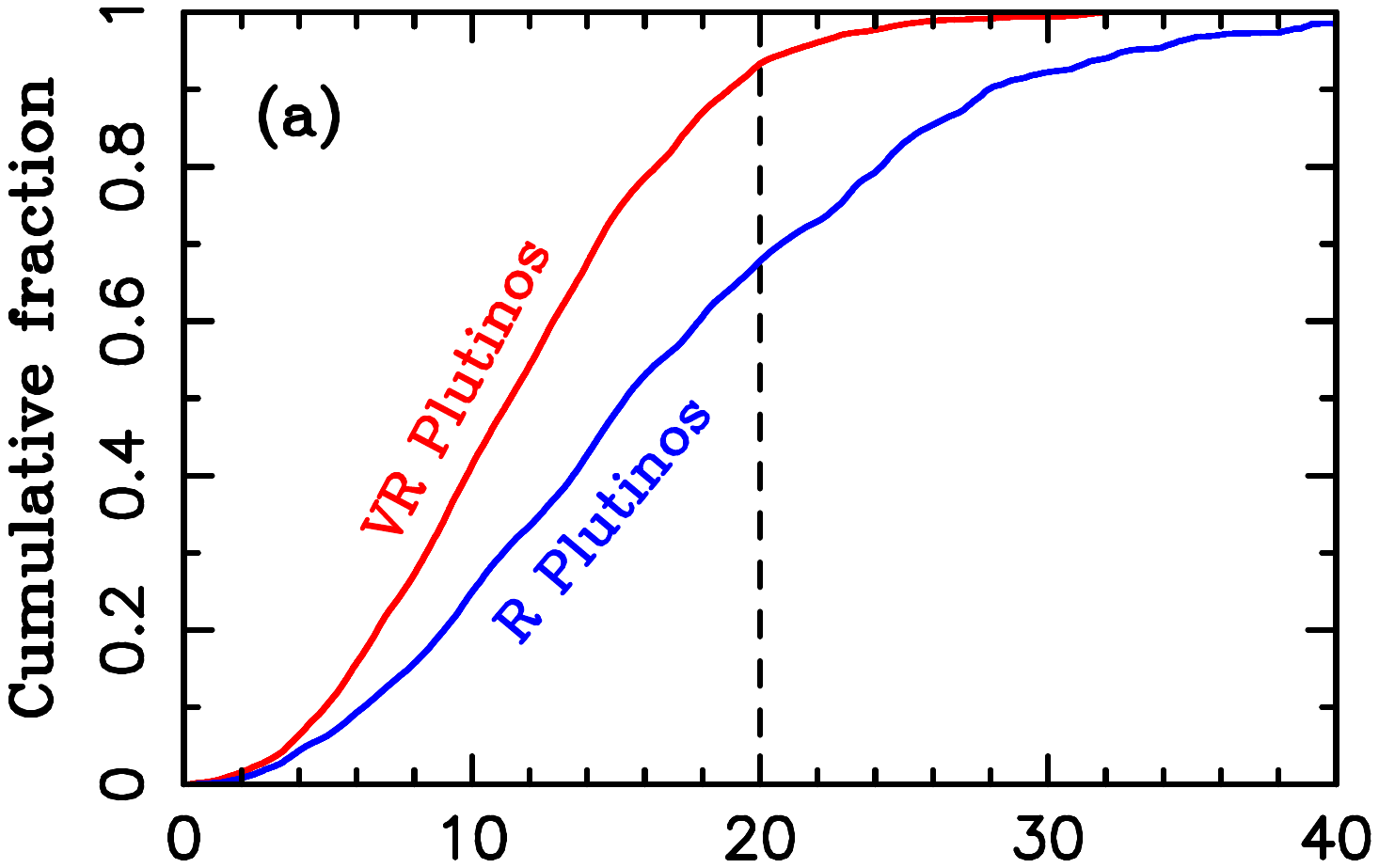}%\\[2.mm]
\plotone{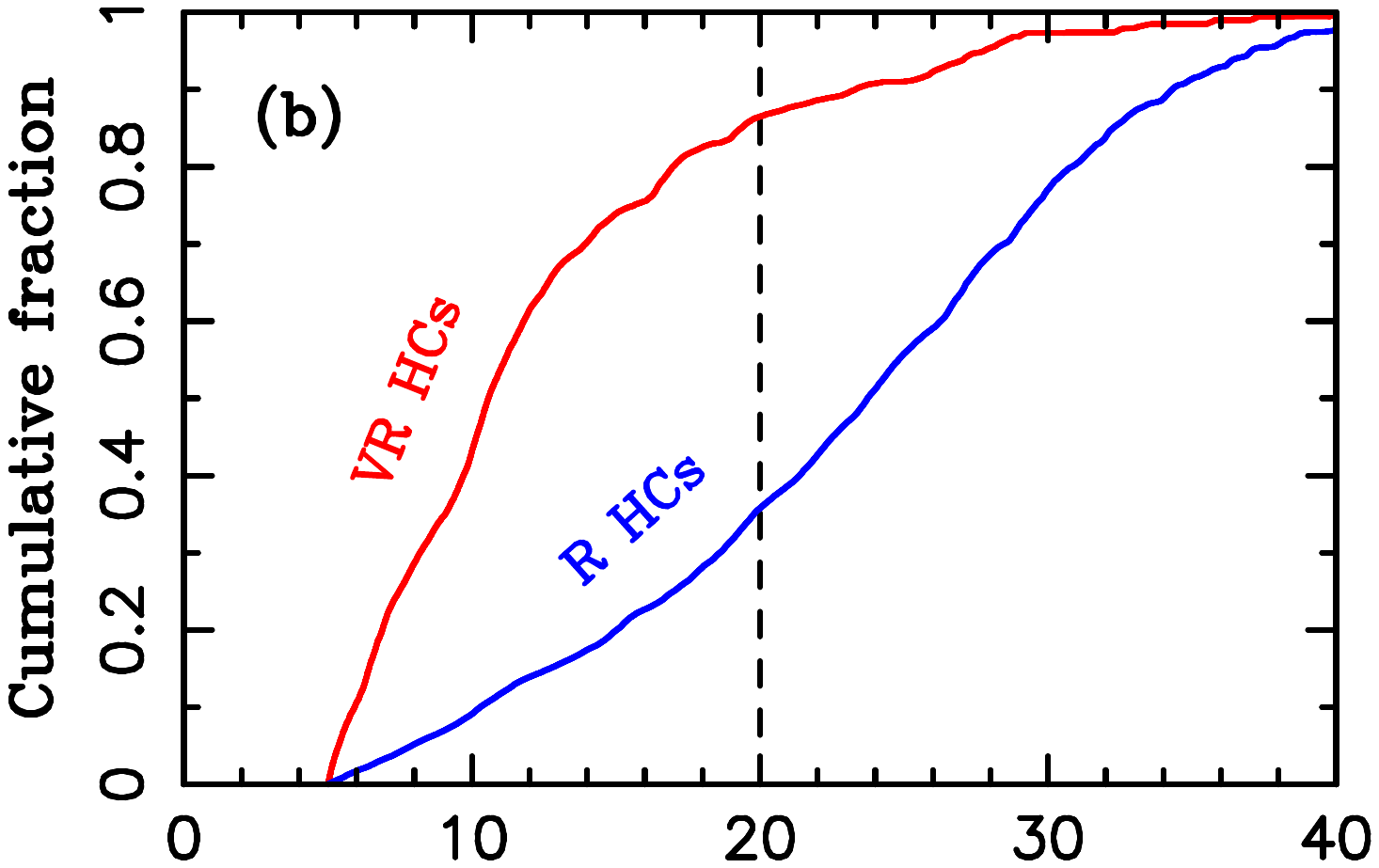}%\\[2.mm]
\plotone{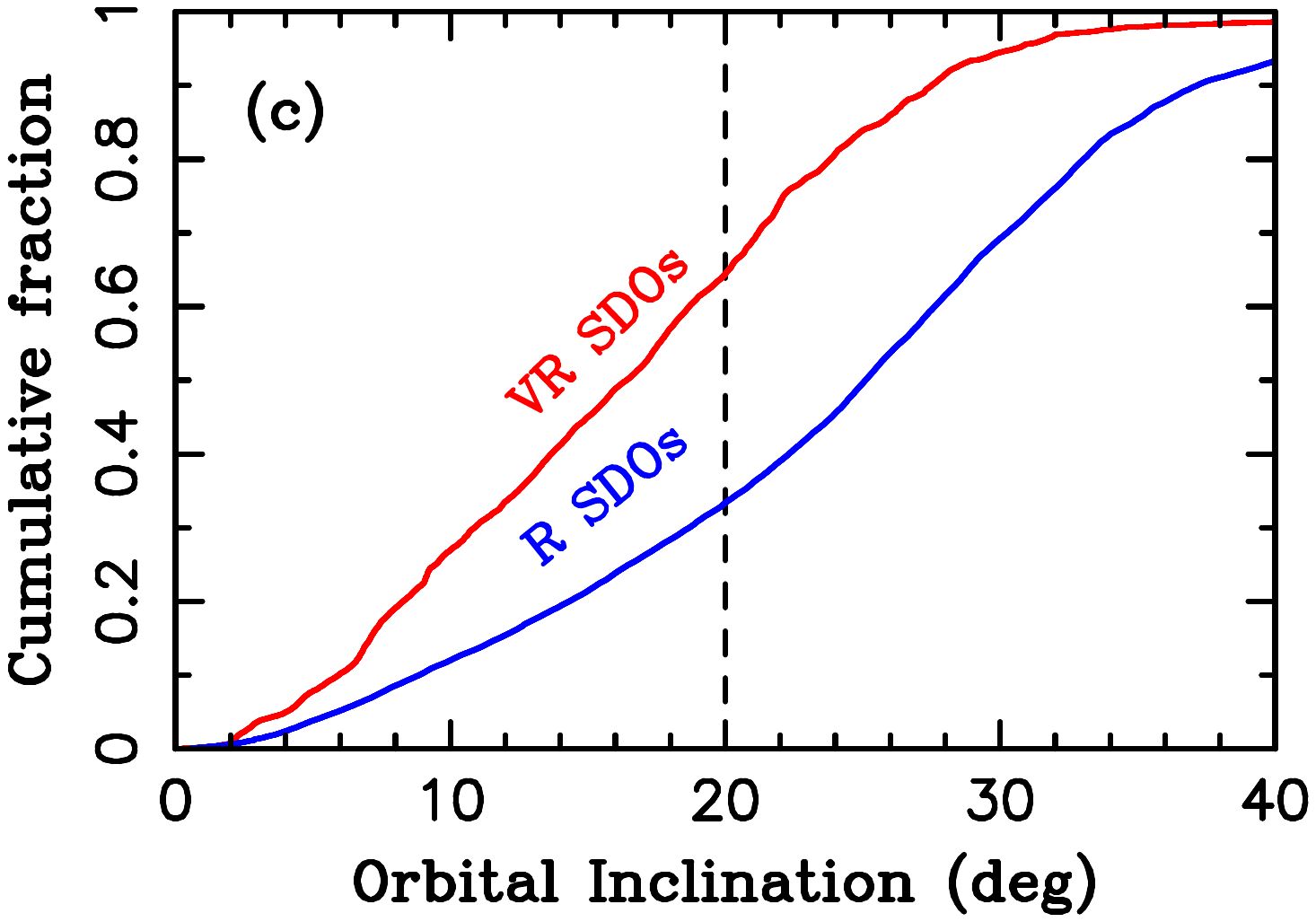}
\caption{The proposed color hypothesis implies that the occurrence of R and VR objects in the hot 
populations should correlate with orbital inclination. Here we take the model results from Fig. 
\ref{marsset} and plot them as cumulative distributions. Panels (a), (b) and (c) show Plutinos,
HCs and SDOs, respectively. In all cases, the R bodies have a significantly broader 
inclination distribution than the VR bodies, as observed.}   
\label{correl}
\end{figure}

Figure \ref{correl} demonstrates this in more detail. Here we re-plot the model results from Fig. \ref{marsset} 
as cumulative distributions. For example, $\simeq$95\% of VR Plutinos 
are expected to have $i<20^\circ$, in a close correspondence to observations (Marsset et al. 2019). In 
contrast, $\simeq$30\% of R Plutinos have $i>20^\circ$ (Fig. \ref{correl}a). The R category dominates in 
the Plutino population, and, when biased, matches the OSSOS inclination distribution in Figure~\ref{plutino}b. 
Both R and VR SDOs are expected to have broader inclination distributions than Plutinos (Fig. 
\ref{correl}c). The inclination distribution of HCs is intermediate between Plutinos and SDOs. 
Interestingly, the HCs also show the biggest difference in the inclination distribution of R and VR 
bodies (Fig. \ref{correl}b). This may be reflected in Fig. 3 of Marsset et al. (2019), where HCs
characterized as VR by OSSOS cluster near $i=10^\circ$.    

\begin{figure}
\epsscale{0.6}
\plotone{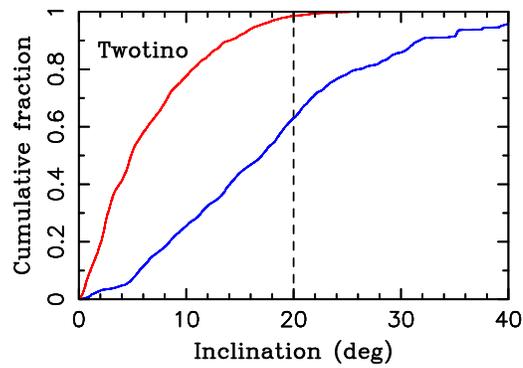}
\caption{The intrinsic inclination distributions of R (blue line) and VR (red line) Twotinos in 
the 2:1 resonance with Neptune obtained in the s30/100j model ($\Delta r=2.5$ au and $r^*=37$ au). 
About 80\% of VR Twotinos are expected to have $i<10^\circ$.}   
\label{twotino}
\end{figure}

Finally, in Fig. \ref{twotino}, we show the inclination distributions of R and VR bodies in the 2:1 
resonance (Twotinos). Overall, we find the intrinsic VR/R ratio to be $\simeq$0.4, whereas Sheppard (2012) 
reported 4 VR and 6 R Twotinos, suggesting observed VR/R $\sim0.67$. These numbers indicate that Twotinos are 
a more equal mix of R and VR objects than other KBO populations (where, typically, one of the color groups 
dominates). In our model, this is caused by the 2:1 resonance migration over the $r<47$ au region and capture 
of low-inclination VR objects into the 2:1 resonance. The orbital inclinations are somewhat increased by 
capture but, as Fig. \ref{twotino} shows, $\sim$80\% of VR objects in the 2:1 resonance still have 
$i<10^\circ$. Our expectation is thus that the low-$i$ Twotinos should be predominantly VR, whereas Sheppard (2012) 
reported 2 R and 1 VR Twotinos with $i<5^\circ$. It would be useful to obtain colors for more low-$i$ Twotinos 
to understand whether they are indeed predominantly VR, as our model suggests. Another consequence of capture 
of low-$i$ objects into the 2:1 resonance is that the inclination distribution of Plutinos should be narrower 
than that of the 3:2 or 5:2 resonant populations. And this is indeed supported by observations (e.g., 
Chen et al. 2019).

\section{Conclusions}

The main findings of this work are:
\begin{enumerate}
\item The dynamical models with slow migration of Neptune reproduce the number of KBOs in different 
populations and their orbital distribution. The migration timescale is inferred to be intermediate 
between the s10/30j and s30/100j models (Table 1). The jumping Neptune model can explain the Kuiper belt kernel.   
\item The different proportions of R and VR colors in different KBO populations can be explained 
if the R bodies formed at radial distances $r<r^*$ and the VR bodies at $r>r^*$, with $30<r^*<40$ au. 
The subsequent evolution mixed R and VR bodies into different populations.
\item The R to VR transition at $30<r^*<40$ au in the original planetesimal disk implies that
the inclination distribution of R bodies should be broader than that of VR bodies, in a close correspondence 
to observations. This results provides support to the color hypothesis proposed here.
\item The exponential ($2.2 < \Delta r < 3.1$ au) and truncated power-law ($300<c<3000$) profiles 
of the original disk work equally well to reproduce various constraints. Additional work will be needed 
to distinguish between these possibilities.  
\end{enumerate}

The suggested transition from R to VR colors at 30-40 au can be a consequence of the 
sublimation-driven surface depletion in some organic molecules, such as NH$_3$ (Brown et al. 2011). 
In this case, the transition should have happened at the sublimation radius. An alternative 
possibility is that the R-to-VR transition traces different collisional histories of objects. 
Consider that the original planetesimal disk below 30 au was massive and could have suffered intense 
collisional grinding over its lifetime ($t_{\rm disk}<100$ Myr; Nesvorn\'y 2018). This 
may have affected the surface properties of the planetesimals that emerged from $<$30~au (via impact related 
depletion and burial of volatiles). 
In contrast, the collisional activity beyond $\sim$30 au should have been relatively modest due to a lower 
disk surface density in this region. Note, however, that the magnitude distributions of R and VR objects 
in the dynamically hot populations are similar, at least in the range $7.5<H<9$ examined by Wong 
\& Brown (2017), thus ruling out a substantial difference in the collisional history of 100-km-class KBOs. 

Centaurs, presumed to have relatively recently evolved from the Kuiper belt, share the bimodality 
of colors and color-inclination correlation with hot KBOs (e.g., Wong \& Brown 2017). Observations show 
that the VR colors of Centaurs disappear when objects reach $r \lesssim 10$ au (Jewitt 2015), probably due 
to the increased heating and removal/burial of the very red matter. This explains why Jupiter 
Trojans at 5 au cannot have VR colors. The primary reason behind the color similarity of Jupiter and 
Neptune Trojans (Jewitt 2018), however, is that both these populations formed at $r<r^*=30$-40 au and did not 
have the VR colors to start with. 

\section{Appendix A}

The break in the size distributions of Jupiter Trojans and dynamically hot KBOs is often interpreted as a result of 
collisional grinding of planetesimals in the massive original disk before its dispersal by Neptune 
(e.g., Nesvorn\'y \& Vokrouhlick\'y 2019; see also Pan \& Sari 2005, Fraser 2009, Campo Bagatin \& Benavidez 
2012). The CC population, instead, given its low mass and distant orbits, has probably not experienced 
any intense period of collisional grinding. Its size distribution may thus reflect the initial mass 
function of planetesimals that was determined by early formation processes.

Here we use, inspired by existing simulations of the streaming instability (Youdin \& Goodman 
2005; see, e.g., Simon et al. 2017, Li et al. 2018), the cumulative size distribution 
\begin{equation}   
N(>\!\!D) = A \left( {D \over D_0} \right)^{-p} 
\exp \left[ - \left( {D \over D_0} \right)^q \right ] \,  
\end{equation}
where $A$, $p$, $q$ and $D_0$ are parameters. Here, $p$ is the cumulative power-law slope index 
of small bodies. The power law is exponentially tampered for bodies with $D \gtrsim D_0$. The albedo 
$p_{\rm V}=0.1$ is assumed to convert sizes to absolute magnitudes. By forward modeling the OSSOS observations 
of CCs, we find $A \simeq 1.3 \times 10^5$, $p\simeq0.6$, $q\simeq1.2$ and $D_0\simeq60$~km.
The OSSOS calibration therefore implies $\sim$15,000 CCs (this is the number used in the main text) with $D>100$ 
km, which is a factor of several smaller population than the one reported by Petit et al. (2011). If 
$p_{\rm V}=0.2$ instead, we obtain $\sim$5,000 CCs with $D>100$ km. For comparison, CCs have mean 
$p_{\rm V}=0.13 \pm 0.05$ (M\"uller et al. 2020) and Arrokoth --member of the CC population-- has $p_{\rm V}=0.23$ 
(Spencer et al. 2020). 

Our forward modeling of OSSOS detections also indicates that the current mass of CCs is $M_{\rm CC} \simeq 5\times10^{-3}$ 
$M_{\rm Earth}$ for $\rho=1$ g cm$^{-3}$ bulk density and $p_{\rm V}=0.1$. If, instead, $p_{\rm V}=0.2$, then 
$M_{\rm CC} \simeq 2\times10^{-3}$ $M_{\rm Earth}$. Adopting $\rho=0.5$ g cm$^{-3}$, as motivated by observations of Arrokoth and 
some CC binaries (Noll et al. 2020), would halve the total mass. In summary, we find $M_{\rm CC} = (3 \pm 2) \times10^{-3}$ 
$M_{\rm Earth}$, roughly $\sim$3-15 larger than the CC mass estimated in Fraser et al. (2014). Note that the size 
distribution is uncertain for $D<D_0$ where OSSOS did not detect a statistically large number of CCs. 

In summary, we use the broken power-law distribution for bodies starting in the massive disk 
below 30 au and the exponentially tampered power-law distribution for bodies starting beyond 40 au. 
The transition radius, $r_{\rm trans}$ between these distributions is uncertain. We tested $30 < 
r_{\rm trans} < 40$ au and found that the orbital and color distribution of $D>100$ km bodies is relatively 
insensitive to the exact location of this transition. The $r_{\rm trans}$ parameter would mainly 
influence the population of very small KBOs, where the two distribution considered here differ the most, 
but that's not the subject of this work. All the results reported in the main text were obtained with 
$r_{\rm trans}=40$ au. The results with no transition (i.e., for $r_{\rm trans}>50$ au) are 
similar.

\acknowledgements

D.N.'s work was supported by the NASA Emerging Worlds program. The work of D.V. was supported 
by the Czech Science Foundation (grant 18-06083S). M.T.B. appreciates support during OSSOS from UK 
STFC grant ST/L000709/1, the National Research Council of Canada, and the National Science and 
Engineering Research Council of Canada. K.V. acknowledges support from NASA grants NNX14AG93G and 
NNX15AH59G.

\clearpage
\begin{table}
\centering
{
\begin{tabular}{lrrrrrrr}
\hline \hline
migration       & $a_{{\rm N},0}$ & $\tau_1$ & $\tau_2$ & $\Delta a_{\rm N}$ & $N_{\rm Pluto}$  & $a_{\rm N}$ & $P_{\rm N}/P_{\rm U}$ \\   
model            & (au)       & (Myr)    & (Myr)        &    (au)    &   & (au) & \\  
\hline
s10/30        & 24         & 10       & 30          &   0               & 2000 & 29.6 & 1.92   \\
s10/30j       & 24         & 10       & 30          &   0.4             & 2000 & 30.1 & 1.95   \\
s30/100       & 24         & 30       & 100         &   0               & 4000 & 29.7 & 1.95   \\
s30/100j      & 24         & 30       & 100         &   0.5             & 4000 & 29.9 & 1.94   \\
\hline \hline
\end{tabular}
}
\caption{A two stage migration of Neptune was adopted from Nesvorn\'y \& Vokrouhlick\'y (2016): 
$\tau_1$ and $\tau_2$ define the $e$-folding exponential migration timescales during these stages,
$a_{\rm N,0}$ and $a_{\rm N}$ denotes Neptune's initial and final semimajor axes, $\Delta a_{\rm N}$
is the jump applied to Neptune's semimajor axis at the transition between stages 1 and 2, 
$N_{\rm Pluto}$ is the assumed initial number of Pluto-mass objects in the massive disk below 30 au
(Nesvorn\'y \& Vokrouhlick\'y 2016). 
The last columns show the final semimajor axis of Neptune and the orbital period ratio of Uranus and Neptune. For reference,
$a_{\rm N}=30.11$ au and $P_{\rm N}/P_{\rm U}=1.96$ in the current Solar System. The leading letter ``s'' 
in the simulation names indicates that both these migrations are considered to be slow (Nesvorn\'y 2015a), the 
trailing letter ``j'' indicates whether a jump has been applied to Neptune at the transition 
between Stages 1 and 2. }
\end{table}

\clearpage

\clearpage

\clearpage

\clearpage

\clearpage

\clearpage

%\clearpage
%\begin{figure}
%\epsscale{0.55}
%\plotone{hots.eps}
%\caption{A comparison between the biased model (solid black line) and OSSOS observations (red dots) 
%of HCs in the 3:2 resonance with Neptune. These results were obtained with an exponential disk profile 
%($r_0=24$ au, $\Delta r = 4$ au, $M_{\rm disk}=20$ $M_{\rm Earth}$) and a broken power-law size distribution 
%from Fraser et al. (2014). See Figure \ref{case40} for Neptune's migration in this case. }
%\label{hots}
%\end{figure}

%\clearpage
%\begin{figure}
%\epsscale{0.55}
%\plotone{scatter.eps}
%\caption{A comparison between the biased model (s10/30j; solid black line) and OSSOS observations (red dots) 
%of SDOs. These results were obtained with an exponential disk profile ($r_0=24$ au, $\Delta r = 2.5$ au, 
%$M_{\rm disk}=20$ $M_{\rm Earth}$) and a broken power-law size distribution from Fraser et al. (2014).}
%\label{scatter}
%\end{figure}

\clearpage

\clearpage

\clearpage

\clearpage

\clearpage

\clearpage

\end{document}